\documentclass[eqsecnum,pra]{revtex4}

\usepackage{amssymb}
\usepackage{amsmath}
\usepackage{rotating}
\usepackage{multirow}
\usepackage{amsfonts}

\begin{document}
\title{Analytic solutions of the 1D finite coupling delta function Bose gas}
\author{P. J. Forrester}
\affiliation{Department of Mathematics and Statistics, The University of Melbourne, Parkville, 3010, Australia}
\author{N. E. Frankel}
\affiliation{School of Physics, The University of Melbourne, Parkville, 3010, Australia}
\author{M. I. Makin}
\affiliation{School of Physics, The University of Melbourne, Parkville, 3010, Australia}

\newcommand{\ii}{{\rm i}}
\newcommand{\manyints}{\int dx_1\ldots\int dx_N}
\newcommand{\detplane}{\det[e^{\ii k_j\tilde{x}_l}]}
\newcommand{\detplanesq}{|\detplane|^2}
\newcommand{\dd}[1]{\frac{\partial}{\partial #1}}
\newcommand{\pairatinf}{g_{N+2}^{(\infty)}(x,0)}
\newcommand{\pairatinfatN}{g_{N}^{(\infty)}(x,0)}
\newcommand{\pairatc}{g_{N+2}(x,0)}
\newcommand{\freddet}[1]{\Delta(x;#1)}
\newcommand{\fredone}{\Delta_1\!\!\left(\begin{array}{l}x\\0\end{array};\lambda\right)}
\newcommand{\fredonem}[3]{\Delta_1\!\!\left(\begin{array}{l}#1\\#2\end{array};#3\right)}
\newcommand{\cinf}[1]{\textrm{O}\left(\frac{1}{cL}\right)^{#1}}
\newcommand{\czero}[1]{\textrm{O}(cL)^{#1}}
\newcommand{\bbs}{\!\!\!\!\!\!}
\newcommand{\ninfinv}{\frac{1}{({\mathcal N}^{(\infty)})^2}}
\newcommand{\sldots}{}

\begin{abstract}
  An intensive study for both the weak coupling and strong coupling
  limits of the ground state properties of this classic system is
  presented.  Detailed results for specific values of finite $N$ are
  given and from them results for general $N$ are determined.  We
  focus on the density matrix and concomitantly its Fourier transform,
  the occupation numbers, along with the pair correlation function and
  concomitantly its Fourier transform, the structure factor.  These
  are the signature quantities of the Bose gas.  One specific result
  is that for weak coupling a rational polynomial structure holds
  despite the transcendental nature of the Bethe equations.  All these
  new results are predicated on the Bethe ansatz and are built upon
  the seminal works of the past.
\end{abstract}
\maketitle
\section{Introduction}

The one dimensional delta function Bose gas is a classic in the field
of exactly solvable integrable systems. Following on the realization
that in the infinite coupling limit, the impenetrable limit, the
system had many of the properties of the free Fermi gas
\cite{ref:girardeau}, in their seminal work, Lieb and Liniger
\cite{ref:liebliniger} solved the model exactly. They derived the
Bethe ansatz and Bethe equations and went on to solve for the
excitations in the thermodynamic limit
\cite{ref:liebliniger,ref:liebII}.

In this paper, we present an intensive study of the finite system, in
the weak coupling and the strong coupling limits.  Extensive
analytical solutions are given for finite values of $N$, the particle
number, and from them, the analytical solutions for general $N$ are
determined. Except for a couple of papers that give the excitations
for a system of three particles \cite{ref:mugasnider} and larger
number of particles \cite{ref:sakmann,ref:batchelorhard} this is the
first intense study of this finite $N$ body system for finite
coupling.  We concentrate on the principle quantities that are the
signatures of the Bose gas.

After a preliminary Section \ref{sec:constructingthewavefunction} and
Appendix \ref{sec:smallc}, introducing the Bethe ansatz and Bethe
equations and their properties, we give extensive analytical
solutions, in Sections \ref{sec:density} and \ref{sec:densitylargecL},
with the aide of Appendices \ref{sec:fredholm} and
\ref{sec:numericaldata}, for the density matrix and concomitantly the
occupation numbers.  In Sections \ref{sec:correlation} and
\ref{sec:correlationlargec}, we do likewise for the pair correlation
function and concomitantly the structure factor. As such we build upon
the seminal work of Lenard \cite{ref:lenard}, the very important work
of Jimbo and Miwa \cite{ref:jimbomiwa}, the Leningrad group
\cite{ref:korepinbook}, and our recent works
\cite{ref:ffgw03,ref:ffg,ref:ffgw03pain}. We conclude the paper in
Section \ref{sec:conclusion}, with some further comments following
upon all these results.

The occupation numbers and structure factors for the modes are the
experimentally realizable signatures of the Bose gas. The spur to the
recent, over the past decade, revival of active interest in this
system is due to Olshanii \cite{ref:olshanii}, who pointed out how
this system could be realized in nature and to the current
experimental activity seeking to realize it \cite{ref:stoferleretal,
  ref:tolraetal,ref:paredesetal,ref:kinoshitawengerweiss,
  ref:kohletal, ref:fertigetal}.  We refer to the introduction in our
earlier paper \cite{ref:ffgw03} for a discussion of the relevant
physical parameters involved.

As all the new analytical solutions in this work are based upon the
Bethe ansatz, it is appropriate to observe that this year celebrates
the 75$^{\rm th}$ anniversary of Bethe's paper \cite{ref:bethe} in
which he introduced the now famous ansatz in his study of the
Heisenberg spin-chain.  Ever since, it has been a golden key in
unlocking the solutions to many exactly solvable integrable systems.
Baxter \cite{ref:baxter} gave a concise review of these in his tribute
to Yang, who has recently written \cite{ref:habandhisphysics} on the
Bethe ansatz to honor Bethe.

\section{Constructing the wavefunction}
\label{sec:constructingthewavefunction}

The purpose of this section is to present the Bethe ansatz
wavefunction and concomitantly the Bethe equations for the
one-dimensional delta function Bose gas in periodic boundary
conditions, as derived in the seminal paper by Lieb and Liniger
\cite{ref:liebliniger}.  The Schr\"{o}dinger equation for this system
with $\hbar=1, 2m=1$ is

\begin{equation}
\label{eq:schro}
\left( - \sum_{i=1}^N \frac{\partial^2}{\partial x_i^2} + 2 c \!\!\!\!\sum_{1\leq i<j\leq N}\bbs\delta(x_i-x_j)\right) \psi_N(x_1,x_2,\ldots,x_N) = E \psi_N(x_1,x_2,\ldots,x_N)
\end{equation}
\\
where $c$ controls the strength of the $\delta$-function.  Throughout
this paper we consider only $c\geq0$, the repulsive case.

Here we display the solution for periodic boundary conditions

\begin{equation}
  \label{eq:wavefunctiondefn}
  \psi_N(x_1,x_2,\ldots,x_N) = \sum_{p\in S_N} a(p) \epsilon(p) e^{\left(\ii \sum_{j=1}^N k_{p(j)} x_j\right)}
\end{equation}
\\
where $S_N$ is the symmetric group on $N$ symbols, and hence the
wavefunction is a sum over $N!$ quantities.  The function
$\epsilon(p)$ is the signature of the permutation $p$, and the function $a(p)$
is given by

\begin{equation}
\label{eq:jimbomiwaa}
a(p) = \prod_{i<j}^N \left[ 1 + \frac{\ii}{c} (k_{p(j)}-k_{p(i)})\right].
\end{equation}

Note that $a(p)$ may be given in a number of forms (any change
absorbed into the normalisation), we choose (\ref{eq:jimbomiwaa}) from
Jimbo and Miwa~\cite{ref:jimbomiwa}.  Throughout this paper we use the
normalisation given by

\begin{equation}
  \label{eq:normalisation}
  {\mathcal N}^2 = \int_{R_{N-1}}\bbs\! dx_1 \ldots dx_{N-1} |\psi_N(0,x_1,\ldots,x_{N-1})|^2
\end{equation}
\\
where $R_{N-1}$ is the domain of integration specified by

\begin{equation}
  \label{eq:RN}
  R_{N-1} : 0 < x_1 < \ldots < x_{N-1} < L.
\end{equation}

In this paper we are concerned only with the ground state of the
system, where $\sum_{i=1}^N k_i=0$, which leads to

\begin{equation}
  k_i=-k_{N+1-i} \quad\forall\; i=1,\ldots,N.
\end{equation}

The $k_i$ are ordered such that $k_N > k_{N-1} > \ldots > k_2 > k_1$.
The $N$ (real) numbers $k_i$ are determined as the solution of the
Bethe equations.  These can be given in many forms
\cite{ref:jimbomiwa,ref:gaudin,ref:liebliniger}, we display that of
\cite{ref:liebII}

\begin{equation}
  \label{eq:BAEourform}
  (k_{j+1}-k_j) L = 2\pi+\sum_{i=1}^N \left[2\arctan\left(\frac{k_i-k_{j+1}}{c}\right) - 2\arctan\left(\frac{k_i-k_j}{c}\right)\right] \quad j=1,\ldots,N-1.
\end{equation}

While these equations cannot be solved explicitly for $k_j$ as a
function of $c$, they can be solved for both small and large $cL$
expansions, using the method of quadrature.  We list explicit small
$cL$ expansions in Appendix \ref{sec:smallc}, and at this point we
highlight that

\begin{equation}
  \label{eq:smallck}
  k_j = \sqrt{\frac{2c}{L}} h_j^{(N)} \left(1- \frac{1}{24}(cL)+\czero{2}\right)
\end{equation}
\\
where $h_j^{(N)}$ is the $j$th zero of the $N$th Hermite polynomial.
The leading term in (\ref{eq:smallck}) is found in reference to this
problem by Gaudin \cite{ref:gaudin}, for more detail refer to
Szeg\"{o} \cite{ref:szego}.  The coefficient $-1/24$ in the next term
of the expansion is new, and appears to be universal (see Appendix
\ref{sec:smallc}).

Here we have chosen to display the large $cL$ expansion for $k_j$ in
its general $N$ form as given in \cite{ref:jimbomiwa}, for its
particular utility in the following sections.

\begin{eqnarray}
  \label{eq:jimbomiwalargeskappa}
  k_j&=&(2j-N-1)\frac{\pi}{L} \Bigg[1-2N\left(\frac{1}{c L}\right)+4N^2\left(\frac{1}{c L}\right)^2\nonumber\\
  &&\qquad\qquad\qquad\qquad+ \left\{-8 N^3+\frac{4}{3} N \left[2 j^2+(N+1) (N-2 j)\right] \pi ^2\right\}\left(\frac{1}{cL}\right)^3\Bigg]+\cinf{4}.
\end{eqnarray}

We also list some exact solutions to the Bethe equations in Table
\ref{tab:BAEexactsolns}.

\begin{table}[htbp]
  \centering
  \begin{tabular}{ccc}
    \hline
    $N$ & $c L$ & $k_N L$ \\
    \hline
    \hline
    2 & $\quad\pi$ & $\quad\pi/2$\\
    3 & $\quad\pi/2$ & $\quad\pi(\sqrt{17}-3)/4$\\
    3 & $\quad\pi$ & $\quad\sqrt{2}\pi$\\
    3 & $\quad3\pi/2$ &$\quad3\pi(\sqrt{17}+3)/4$ \\
  \end{tabular}
  \caption{A sample of exact solutions to (\ref{eq:BAEourform})}
  \label{tab:BAEexactsolns}
\end{table}

It is now possible to explicitly construct the wavefunction
$\psi_N(x_1,\ldots,x_N)$ using (\ref{eq:wavefunctiondefn}) and
(\ref{eq:jimbomiwaa}).  We close here by exhibiting the unnormalised
wavefunction $\psi_2(x_1,x_2)$ by way of example

\begin{equation}
  \psi_2(x_1,x_2) = \left(1-\frac{2\ii k_2 }{c} \right) e^{\ii k_2 (x_2-x_1)}-\left(1+\frac{2\ii k_2}{c} \right) e^{-\ii k_2 (x_2-x_1)}.
\end{equation}

which has normalisation

\begin{equation} 
  {\mathcal N}^2=2 L-\frac{\sin2k_2L}{k_2}-\frac{8 \sin
    ^2k_2 L}{c}+\frac{4 k_2 (2 k_2 L+\sin2k_2L)}{c^2}.
\end{equation}

In all that follows the weak coupling expansion corresponds to the
dimensionless parameter $cL \ll 1$ and the strong coupling expansion corresponds to $cL \gg 1$.

\section{Density Matrix and Occupation Numbers}
\label{sec:density}

The normalised density matrix $\rho_N(x,0)$ is defined as 

\begin{equation}
  \label{eq:densitydefn}
  \rho_N(x,0) = \frac{N}{L}\frac{1}{\mathcal N^2} \sum_{j=0}^{N-1}\int_{R_{N-1,j}(x)}\bbs\bbs\bbs dx_1 \ldots  dx_{N-1} \psi_N(0,x_1,x_2,\ldots,x_{N-1})\overline{\psi_N(x_1,\ldots,x_j,x,x_{j+1},\ldots,x_{N-1})} 
\end{equation}
\\
where the overbar implies complex conjugation, the normalisation
${\mathcal N}^2$ is specified by (\ref{eq:normalisation}), and the
domain of integration is specified by

\begin{equation}
  \label{eq:RNj}
  R_{N,j}(x):0\leq x_1 < \ldots < x_j < x < x_{j+1} < \ldots < x_N \leq L.
\end{equation}

The density matrix is normalised such that $\rho_N(0,0)=\rho_0=N/L$.
Hence to compute the density matrix for $N$ particles, one must
perform $N$ lots of $N-1$ dimensional integrals over $(N!)^2$ terms
(e.g.~4 triple integrals over 576 terms for $N=4$).  This is a
computationally expensive task, and hence we were able to determine
solutions only up to $N=4$ using this method.

The occupation numbers $c_n(N)$ are determined as a Fourier transform
of the density matrix

\begin{equation}
  \label{eq:occupationdefn}
  c_n(N)=\int_0^L \rho_N(x,0) e^{2\ii \pi n x/L}dx =\int_0^L \rho_N(x,0) \cos(2\pi n x/L)dx.
\end{equation}
\\
They have the physical interpretation of being the expectation value
of the number of particles in mode $n$.

We note the normalisation property $\sum_{n=-\infty}^{\infty}
c_n(N)=N$; we have confirmed this result for all occupation number
formulas that follow. It is sometimes useful to discuss occupation
number per particle; we introduce the notation $c^*_n(N)=c_n(N)/N$.
We shall now display some explicit solutions for $N=2,3,4$ and general
$N$.

\subsection{$N=2$}

Within Section \ref{sec:density}, this subsection heralds the most
complete set of results, with results for $N=3$ and 4 becoming
increasingly exiguous as the intricacies of the equations develop.
For example, it is only possible to display the complete density
matrix for $N=2$, as for $N=3$ already the equation would take many
pages to display.  We have utilised for $N=2$ (\ref{eq:BAEourform}) to
obtain

\begin{equation}
  \label{eq:BAENis2exact}
  c = 2 k_2  \tan\left(\frac{k_2 L}{2}\right)
\end{equation}
\\
hence producing a concise form of the density matrix using
(\ref{eq:wavefunctiondefn}), (\ref{eq:jimbomiwaa}) and
(\ref{eq:densitydefn})

\begin{equation}
  \label{eq:density2general}
  \qquad\qquad\rho_2(x,0)= \frac{2}{L}\frac{k_2x\cos (k_2\left( L - x \right) ) + 
    k_2\left( L - x \right) \cos (k_2x) + 
    \sin (k_2\left( L - x \right) ) + \sin (k_2x)}{k_2L + \sin (k_2L)} \qquad 0 \leq k_2 \leq \pi
\end{equation}
\\
with corresponding occupation numbers from (\ref{eq:occupationdefn})

\begin{equation}
  \label{eq:smallcOccNis2}
  \quad c_n(2)=2\frac{4(k_2 L)^3\left( 1 - \cos (k_2L) \right) }
  {(4n^2\pi^2-k_2^2L^2)^2\left( k_2L + \sin (k_2L) \right) } \qquad 0 \leq k_2 \leq \pi.
\end{equation}

The expansion of the density matrix for small $cL$ is given using
(\ref{eq:BAENis2})

\begin{eqnarray}
  \label{eq:smallcdensity2}
  \rho_2(t,0)&=& \frac{2}{L}\Big[1 - \frac{t^2 (\pi-t)^2}{24 \pi^4} (c L)^2+\frac{t^2 (\pi-t)^2 (t^2-\pi t + 2\pi^2)}{360\pi^6} (c L)^3\nonumber\\
  &&+\frac{t^2(\pi-t)^2
    \left( 16{\pi }^4 - 24{\pi }^3t + 27{\pi }^2t^2 -  6\pi t^3 + 3t^4 \right) }{40320{\pi }^8}(c L)^4+\czero{5}\Big].
\end{eqnarray}

Note that we have introduced here $t=\pi x/L$, henceforth we switch
between $t$ and $x$ as appropriate.  The corresponding occupation
numbers for (\ref{eq:smallcdensity2}) are given by

\begin{equation}
  c_n(2) = 2\left\{ \begin{array}{ll}
      1-\frac{1}{720} (cL)^2+\frac{1}{6048} (cL)^3-\frac{11}{1209600}(cL)^4+\czero{5} & \textrm{when } n=0\vspace{0.1cm}\\
      \frac{1 }{16 n^4 \pi ^4} (cL)^2+ \frac{3-n^2\pi^2}{96n^6\pi^6} (cL)^3 +\frac{4 n^4\pi ^4 -30n^2 \pi ^2 +45}{3840 n^8 \pi ^8}(cL)^4+\czero{5}& \textrm{when } n\neq 0.
    \end{array}\right.
\end{equation}

Utilising (\ref{eq:jimbomiwalargeskappa}), we obtain a large $cL$
expansion for the density matrix

\begin{eqnarray}
  \label{eq:density2largec}
  \rho_2(t,0)&=&\frac{2}{L}\Big\{\frac{( \pi-2t) \cos t + 2\sin t}{\pi}+\frac{8( \pi  - t ) t\sin t}{\pi  }\left(\frac{1}{cL}\right)\nonumber\\
  &&+\frac{8\left[ ( \pi^2 - 3\pi t + 2t^2 )t \cos t - ( \pi^2 + 6\pi t - 6t^2 ) \sin t \right] }{\pi  }\left(\frac{1}{cL}\right)^2+ \cinf{3}\Big\}.
\end{eqnarray}

Note that in the limit $cL\to\infty$, we recover (20) from
\cite{ref:ffgw03}. The corresponding occupation numbers for
(\ref{eq:density2largec}) are given by

\begin{eqnarray}
  \label{eq:occNum2largec}
  c_n(2)&=&2\Bigg[\frac{8}{{(4n^2-1) }^2{\pi }^2} - 
  \frac{32(12n^2+1) }{(4n^2-1)^3\pi^2}\left(\frac{1}{cL}\right)\nonumber\\
  && \quad\quad+  \left\{\frac{3072 n^2\left(4 n^2+1\right)}{\left(4 n^2-1\right)^4
      \pi ^2}-\frac{64 \left(6 n^2-1\right)}{\left(4
        n^2-1\right)^2}\right\}\left(\frac{1}{cL}\right)^2 + \cinf{3}\Bigg]
\end{eqnarray}
\\
which in the limit $cL\to\infty$ recovers (42) from \cite{ref:ffgw03}.
We also display here the density matrix for the exact solution to
(\ref{eq:BAEourform}) from Table \ref{tab:BAEexactsolns}, when
$cL=\pi, k_2L=\pi/2$

\begin{equation}
  \rho_2(t,0)=\frac{2}{L}\frac{(-t+\pi +2) \cos \frac{t}{2}+(t+2) \sin
    \frac{t}{2}}{\pi+2 }
\end{equation}
\\
and the corresponding occupation numbers are given by

\begin{equation}
  \label{eq:smallcOccNis2specific}
  c_n(2)=2\frac{16}{\left(16 n^2-1\right)^2 \pi  (\pi+2 )}.
\end{equation}

\subsection{$N=3$}
\label{sec:density3}

The small $cL$ expansion of the density matrix is given here, using
(\ref{eq:densitydefn}) and (\ref{eq:BAENis3}).

\begin{eqnarray}
  \label{eq:smallcdensity3}
  \rho_3(t,0)&=&\frac{3}{L}\Big[ 1-\frac{t^2(\pi -t)^2}  {12 \pi
    ^4}(c L)^2 - \frac{t^2 (\pi-t)^2(t^2-\pi t-3\pi^2) }{180 \pi
    ^6}(c L)^3\nonumber\\
  &&+\frac{t^2(\pi -t)^2  \left(-101 t^4+202 \pi  t^3-125  \pi ^2 t^2+24 \pi ^3 t+54 \pi ^4\right)}{20160 \pi^8}(cL)^4+\czero{5}\Big].
\end{eqnarray}

Note that the coefficient of the $(cL)^p$ term is a polynomial in $t$
of order $2p$, for $p\geq2$.  This structure is also repeated in
(\ref{eq:smallcdensity2}).

The corresponding occupation numbers for (\ref{eq:smallcdensity3}) are
given by

\begin{equation}
  \label{eq:smallcOccNis3}
  c_n(3) = 3\left\{\begin{array}{ll}
      1-\frac{1}{360} (c L)^2 + \frac{1}{1680}(c L)^3 -\frac{163}{1814400} (cL)^4+ \czero{5}&\textrm{when } n=0\vspace{0.1cm}\\
      \frac{1}{8n^4\pi^4} (c L)^2-\frac{n^2\pi^2+3}{48n^6\pi^6} (c L)^3 +\frac{6 n^4\pi^4 +170n^2 \pi ^2 -1515}{1920 n^8 \pi^8}(c L)^4+\czero{5}&\textrm{when } n\neq 0.
    \end{array}\right.
\end{equation}

We give here the large $cL$ expansion of the density matrix, using
(\ref{eq:densitydefn}) and (\ref{eq:BAEourform})

\begin{eqnarray}
  \label{eq:density3largec}
  \rho_3(t,0) &=& \frac{3}{L}\Big\{\frac{2(\pi-2t)^2+12(\pi-2t) \sin 2t  + 4 (2t-\pi+2)(2t-\pi-2) \cos 2t+\cos 4t + 15}{6 \pi ^2}\nonumber\\
  &+& \frac{4\sin t\left[ \left( \pi  - 2t \right) 
      \left( 1 + 8\pi t - 8t^2 \right) \cos t - 
      \left( \pi  - 2t \right) \cos 3t + 
      8\left( \pi  - t \right) t\sin t \right]}{\pi^2 }\left(\frac{1}{cL}\right) +\cinf{2}\Big\}
\end{eqnarray}
\\
which in the limit $cL\to\infty$ recovers (21) of \cite{ref:ffgw03}.

The occupation numbers corresponding to (\ref{eq:density3largec}) are
given by

\begin{equation}
  \label{eq:occNum3largec}
  c_n(3)=3\left\{\begin{array}{ll}
      \left(\frac{1}{9}+\frac{35}{6\pi^2}\right)+\left(\frac{8}{3}+\frac{35}{\pi^2}\right)\left(\frac{1}{cL}\right)+\cinf{2}&\textrm{when } n=0\vspace{0.1cm}\\
      \frac{1}{9} - \left(\frac{4}{3}+\frac{35}{6\pi^2}\right)\left(\frac{1}{cL}\right)+\cinf{2}&\textrm{when } |n|=1\vspace{0.1cm}\\
      \frac{35}{108\pi^2}-\frac{385}{36\pi^2} \left(\frac{1}{cL}\right)+\cinf{2}&\textrm{when }|n|=2\vspace{0.1cm}\\
      \frac{2 (3n^2+1)}{3n^2(n^2-1)^2 \pi^2}-\frac{4\left( 9n^6 - 28n^4 - 61n^2 +8  \right) }
      {n^2(n^2-1)^3(n^2-4)\pi^2}\left(\frac{1}{c L}\right)+\cinf{2} &\textrm{when } |n|\geq 3.
    \end{array}
  \right.
\end{equation}
\\
which in the limit $cL\to\infty$ recovers (45) of \cite{ref:ffgw03}.
We also display here the density matrix for the exact solution to
(\ref{eq:BAEourform}) from Table \ref{tab:BAEexactsolns}, when $c
L=\sqrt{2}\pi, k_3L=\pi$

\begin{equation}
  \label{eq:density3special}
  \rho_3(t,0)= \frac{3}{L}\Big\{ \frac{8 t^2-8 \pi  t+2 \phi
    (\pi -2 t) \cos t-4 \cos2t+4 \left[\pi 
      \left(2 \sqrt{2} t+5\right)-2 \sqrt{2}
      \left(t^2-5\right)\right] \sin t+\phi \pi +52}{48+3\phi\pi}\Big\}
\end{equation}
\\
where $\phi=10\sqrt{2}+3\pi$.  The corresponding occupation numbers
for (\ref{eq:density3special}) are given by

\begin{equation}
  c_n(3) = 3\left\{\begin{array}{ll}
      \frac{5}{27} + \frac{4{\sqrt{2}}}{\pi } - 
      \frac{4(299 + 71\sqrt{2}\pi) }
      {27(16 + \phi\pi) }&\textrm{when } n=0\vspace{0.1cm}\\
      \frac{64 \sqrt{2}+54 \pi }{81\pi(16  +\phi\pi)} &\textrm{when } |n|=1\vspace{0.1cm}\\
      \frac{4\left( -\pi  + 
          32n^4\left( 2{\sqrt{2}} + \pi  \right)  - 
          4n^2\left( 12{\sqrt{2}} + \pi  \right)  \right)
      }{3n^2\pi{(4n^2-1) }^3(16 + \phi\pi) 
      }&\textrm{when } |n|\geq 2
    \end{array}
  \right.
\end{equation}

The density matrix and occupation numbers for the other exact values
listed in Table \ref{tab:BAEexactsolns} for $N=3$, although
calculated, are too lengthy to display here.

\subsection{$N=4$}

The intricacy of the density matrix at this value of $N$ is already so
great that we go on to calculate the occupation numbers without
explicitly exhibiting it, and as such we give only $c_0(4)$.  We
utilise (\ref{eq:occupationdefn}) and (\ref{eq:BAENis4}) to produce
the result

\begin{equation}
  \label{eq:smallcOccNis4}
  c_0(4)=4\Big[1 - \frac{1}{240}(cL)^2 + \frac{13}{10080}(cL)^3 - \frac{383}{1209600}(cL)^4+\czero{5}\Big].
\end{equation}

\subsection{General $N$}
\label{sec:densitygeneralN}

It is interesting to observe that with the presence of the irrational
numbers in the Bethe equations for small $cL$ (Appendix
\ref{sec:smallc}), that when the final results for the occupation
numbers appear they contain purely rational numbers.  Encouraged by
this remarkable observation and other indications in the preceding
subsections, we looked for a pattern in $N$ for the coefficients in
the small $cL$ expansion for a general $c_0(N)$.  Upon close
examination of (\ref{eq:smallcOccNis2}), (\ref{eq:smallcOccNis3}) and
(\ref{eq:smallcOccNis4}), and with some good fortune, we found the
following polynomial structure for the small $cL$ expansion of the
$n=0$ occupation number for general $N$.

\begin{eqnarray}
  \label{eq:generalc_0(N)}
  c^*_0(N) &=& 1-\frac{N-1}{720} (cL)^2+\frac{N-1}{720} \left( \frac{4(N-1)+1}{42}\right) (c L)^3\nonumber\\
  &&-\frac{N-1}{720} \frac{1}{42} \left(\frac{45 (N-1)^2-5(N-1)-7}{120}\right) (cL)^4 + \czero{5}.
\end{eqnarray}

We have written (\ref{eq:generalc_0(N)}) as given to emphasise the
detailed structure of the coefficients in this expansion.  Note that
the coefficient of $(N-1)(cL)^p$ is a polynomial of order $p-2$ in
$N-1$ for $p\geq 2$.  Therefore to obtain say the coefficient of the
$(cL)^5$ term we would need $c_0(N)$ for four specific values of $N$
in their rational form.  Unfortunately, it was too computationally
expensive for us to obtain any $N-1$ polynomial for any higher order
than $(cL)^4$.  We conjecture that this is the pattern for all $N\geq
2$.

\newcommand{\smallGamma}{\scriptstyle\Gamma}

The fact that the coefficient of $(N-1)(cL)^p$ appears to be a
polynomial of degree $p-2$ in $N-1$ suggests that for any finite $N$
there is always an interval $cL \in [0,D_N)$ such that the series is
convergent, but with $D_N \to 0$ as $N \to \infty$. To quantify this
last point, note that in the thermodynamic limit the dimensionless
parameter is $c/\rho_0 =\; \smallGamma$ and the coefficient is
proportional to $N^{2p-1}$, which suggests that the corresponding
radius of convergence is proportional to $1/N^2$. In particular this
means that no information can be gleaned as to the functional form of
$c_0^*(N)$ as a function of $\smallGamma$ about $\smallGamma = 0$
except that it is not analytic.

\section{Density Matrices and Occupation Numbers for large $cL$}
\label{sec:densitylargecL}

In the previous section, we gave large $cL$ expansions for the density
matrices and occupation numbers for $N=2$ and $N=3$.  Again, it was
numerically prohibitive to go beyond $N=4$, furthermore as can be seen
from the coefficients in these expansions (and as can be witnessed in
the Tables \ref{tab:nis0majorsplit}-\ref{tab:nis2minorsplit} which we
will refer to in what follows), that the numbers are highly irrational
with no hope of finding an analogous pattern as we were fortunate
enough to do in (\ref{eq:generalc_0(N)}).

We therefore turn to a totally different mathematical strategy to
obtain a large $cL$ expansion for these quantities.  In the
impenetrable limit ($cL=\infty$, arbitrary $L$), Lenard
\cite{ref:lenard} developed the theory for the density matrix for
arbitrary $N$, and went on to show that for asymptotically large $N$
that $c_0(N)\sim N^{1/2}$.

In our recent work \cite{ref:ffgw03} we employed Lenard's theory to
obtain the results for the occupation numbers for a range of finite
$N$, and from them determine the results for general $N$, which
continue on to the asymptotically large $N$ limit \footnote{See
  \cite{ref:ffgw03} for a comprehensive list of references for the
  impenetrable limit.}.

To go beyond the impenetrable limit for these quantities, we turn to
the very valuable work of Jimbo and Miwa \cite{ref:jimbomiwa}.
Building upon the work of Lenard \cite{ref:lenard}, they developed an
expansion for the density matrix in the large $cL$ limit for general
$N$ in principle.  We say in principle because while their expansion
is superb in the form given, it is as numerically prohibitive to use
as was the method we employed in the previous section.

To resolve this difficulty we have recast their theory using
mathematical techniques evidenced in our recent work
\cite{ref:ffgw03pain} and presented in great detail by Forrester
\cite{ref:forresterbook} into a new form that is readily amenable to
numerical calculation.  We will find in what follows that the specific
results for $N=2$ and $N=3$ as in Section \ref{sec:density} are useful
specific checks to the theory.

In Subsection \ref{sec:toeplitz} along with Appendix
\ref{sec:fredholm}, we present the full details of our derivation for
the the density matrix.  Following upon that, in Subsection
\ref{sec:occNumslargec} we are now able to calculate the occupation
numbers for a finite range of $N$ values and from these results are
able to determine the results for general $N$ which again continue to
asymptotically large $N$.

\subsection{Toeplitz Determinants}
\label{sec:toeplitz}

The fact that $|\psi_N|^2$ consists of $(N!)^2$ terms means any method
based on term-by-term integration must necessarily be restricted to
small $N$.  To overcome this one must seek out structure in the form
of $\psi_N$, and this structure must be used to reduce the
computational expense required to compute $\rho_N(x,0)$.  Certainly in
the limit $cL\to\infty$ there is structure in
(\ref{eq:wavefunctiondefn}) for then (\ref{eq:jimbomiwalargeskappa})
gives $k_j = (2j-N-1)\pi/L$ while (\ref{eq:jimbomiwaa}) gives
$a(p)=1$, and so

\begin{eqnarray}
  \left.\psi_N(x_1,\ldots,x_N)\right|_{cL\to\infty} &=& \sum_{p\in S_{N}} \epsilon(p) \prod_{j=1}^N e^{\ii\pi(2j-N-1)x_{p(j)}/L}\\
  &=& \det[e^{\ii\pi(2k-N-1)x_j/L}]_{j,k=1,\ldots,N}.
\end{eqnarray}

While the sum consists of $N!$ terms, the determinant can be computed
in O$(N^3)$ arithmetic operations.  Moreover the corresponding density
matrix can also be expressed as a determinant (Lenard
\cite{ref:lenard})

\begin{equation}
  \label{eq:densityinffred}
  \rho_N^{(0)}(x,0) = -\frac{1}{2}\fredonem{x}{0}{-2}
\end{equation}
\\
where

\begin{equation}
  \label{eq:fredonedef}
  \fredone = \lambda e^{-\ii\pi(N-1) x/L}\det\left[A_1(j-k)\right]_{j,k=1,\ldots,N-1}
\end{equation}

\begin{eqnarray}
  \label{eq:A1defgeneral}
  A_1(j-k)&=&\frac{1}{L}\left(\int_0^L+\lambda \int_0^x\right) du (e^{2\pi \ii u/L}-e^{2\pi \ii x/L})(e^{-2\pi \ii u /L} - 1)e^{2\pi \ii u (j-k)/L}\\
  \label{eq:A1defexplicit}&=& \left\{ \begin{array}{ll}
      \frac{\lambda}{2\pi} \sin2t-\frac{t\lambda}{\pi}-1 & \textrm{when } j-k=-1\vspace{0.1cm}\\
      2 e^{\ii t} \left[(\frac{t\lambda}{\pi}+1) \cos t-\frac{\lambda}{\pi} \sin t\right]  & \textrm{when } j-k=0\vspace{0.1cm}\\
      e^{2 \ii t}\left[\frac{\lambda}{2\pi} \sin 2t -   \frac{t \lambda}{\pi} - 1\right] & \textrm{when } j-k=1\vspace{0.1cm}\\
      \frac{2 \lambda \ii e^{\ii(j-k+1) t}}{(j-k) ((j-k)^2-1)\pi} \left[(j-k) \cos((j-k)t) \sin t - \cos t \sin((j-k)t)\right] & \textrm{when } |j-k|\geq2
    \end{array}\right.
\end{eqnarray}
\\
and thus also can be computed in O($N^3$) operations.  In
\cite{ref:ffgw03} this determinant formula was used to compute the
density matrix and the corresponding ground state occupation numbers
up to $N=7$.  It has been shown by Jimbo and Miwa \cite{ref:jimbomiwa}
(see also Section \ref{sec:correlationlargec} below) that determinant
structures persist if $\psi_N$ is expanded in large $cL$, and that
this implies special structures for the expansion of the density
matrix.  In particular, writing

\begin{equation}
  \rho_N(x,0) = \rho_N^{(0)}(x,0)+\left(\frac{1}{cL}\right)\rho_N^{(1)}(x,0)+\cinf{2}
\end{equation}
\\
it was shown in \cite{ref:jimbomiwa} that

\begin{equation}
  \rho_N^{(1)}(x,0)=-2\rho_0 x \dd{x} \rho_N^{(0)}(x,0) + F_N(x)
\end{equation}
\\
where $\rho_N^{(0)}(x,0)$ is specified by (\ref{eq:densityinffred})
and

\begin{equation}
  \label{eq:F_N(x)}
  F_N(x) = \frac{1}{\freddet{-2}}\left.\left[\dd{x}\fredone\dd{\lambda}\freddet{\lambda}+\dd{\lambda}\fredone\dd{x}\freddet{\lambda}-\fredone\frac{\partial^2}{\partial x\partial \lambda} \freddet{\lambda}\right]\right|_{\lambda=-2}.
\end{equation}

Introducing the kernel function

\begin{eqnarray}
  K_{NL}(x,y) &=& \frac{1}{L} \sum_{j=1}^N e^{-\ii\pi (2j-N-1)(x-y)/L}\\
  &=& \frac{\sin [N\pi(x-y)/L]}{L\sin[\pi(x-y)/L]}
\end{eqnarray}
\\
the quantity $\fredonem{x}{y}{\lambda}$ is defined in
\cite{ref:jimbomiwa} as the Fredholm minor

\begin{eqnarray}
  \label{eq:fredonebigdef}
  \fredonem{x}{y}{\lambda} &=& \sum_{l=0}^{\infty} \frac{\lambda^{l+1}}{l!} \int_0^x du_1 \ldots \int_0^x du_l \nonumber\\
  && \times \det\left[
    \begin{array}{ll}
      K_{NL}(x,y) & \left[K_{NL}(x,u_k)\right]_{k=1,\ldots,l}\\
      \left[K_{NL}(u_j,y)\right]_{j=1,\ldots,l}& \left[K_{NL}(u_j,u_k)\right]_{j,k=1,\ldots,l}
    \end{array}
  \right]
\end{eqnarray}
\\
while $\freddet{\lambda}$ is specified as the Fredholm determinant

\begin{equation}
  \label{eq:freddetbigdef}
  \freddet{\lambda} = \sum_{l=0}^{\infty} \frac{\lambda^l}{l!} \int_0^x du_1 \ldots \int_0^x du_l \det\left[K_{NL}(u_j,u_k)\right]_{j,k=1,\ldots,l}.
\end{equation}

Neither (\ref{eq:fredonebigdef}) or (\ref{eq:freddetbigdef}) are
suitable for computation.  However for $\Delta_1$ we have the
determinant formula (\ref{eq:fredonedef}), and $\Delta$ too can be
expressed as a determinant.  This can be seen by developing Lenard's
\cite{ref:lenard} derivation of (\ref{eq:densityinffred}), which
proceeds by writing the summation of determinants of
(\ref{eq:fredonebigdef}) and (\ref{eq:freddetbigdef}) as
multidimensional integrals of a type which can be recognised as
Toeplitz determinants (see Appendix \ref{sec:fredholm}).  In addition
to (\ref{eq:fredonedef}) - (\ref{eq:A1defexplicit}), one deduces that

\begin{equation}
  \freddet{\lambda} = \det\left[A_0(j-k)\right]_{j,k=1,\ldots,N}
\end{equation}
\\
where

\begin{eqnarray}
  A_0(j-k)&=&\frac{1}{L} \left(\int_0^L + \lambda \int_0^x\right)du \exp[2\pi \ii u (j-k)/L]\\
  &=& \left\{\begin{array}{ll}
      1+\frac{\lambda x}{L} & \textrm{when } j-k=0\vspace{0.1cm}\\
      \frac{\lambda}{2 \ii \pi(j-k)} (e^{2 \ii\pi (j-k)x/L}-1) & \textrm{when } j-k\neq 0.
    \end{array}
\right.
\end{eqnarray}

With these determinant formulas $F_N(x)$ is expressed in a computable
form.  This concludes the method necessary to construct the density
matrix expanded in large $cL$.  We now examine the occupation numbers.

\subsection{Occupation Numbers for Large $cL$}
\label{sec:occNumslargec}

The following notation for the occupation number $c_n(N)$ expanded in
large $cL$ is given as

\begin{equation}
  c_n(N)=c_n^{(0)}(N)+\left(\frac{1}{cL}\right) c_n^{(1)}(N)+\cinf{2}.
\end{equation}

We list some exact values of $c_n^{(0)}(N)$ and $c_n^{(1)}(N)$ for
$n=0$ ($N=2,\ldots,7$) and $n=1,2$ ($N=2,\ldots,6$) in Tables
\ref{tab:nis0majorsplit}, \ref{tab:nis1majorsplit} and
\ref{tab:nis2majorsplit}.  The $c_n^{(1)}(N)$ term can be expanded
further as

\begin{eqnarray}
  c_n^{(1)}(N) &=& \int_0^L \left(-2N x\frac{\partial\rho_N^{(0)}(x,0)}{\partial x}+F_N(x)\right) \cos\left(\frac{2\pi n x}{L}\right) dx\\
  &=& c_n^{(1,1)}(N) + c_n^{(1,2)}(N)
\end{eqnarray}
\\
where $F_N(x)$ may be computed using the expressions in Subsection
\ref{sec:toeplitz} above.  The $c_n^{(1,1)}(N)$ term can be simplified
using integration by parts to

\begin{equation}
  c_n^{(1,1)}(N)=-2N^2+2N c_n^{(0)}(N)-4 \pi n N \int_0^L \frac{x}{L} \sin(2\pi n x/L) \rho_n^{(0)}(x) dx .
\end{equation}

We list some exact values of $c_n^{(1,1)}(N)$ and $c_n^{(1,2)}(N)$ for
$n=0,1,2$ with $N=2,\ldots,6$ in Tables \ref{tab:nis0minorsplit},
\ref{tab:nis1minorsplit} and \ref{tab:nis2minorsplit}.  We present all
the numerical values that we computed for $c_n^{*(0)}(N)$,
$c_n^{*(1,1)}(N)$, $c_n^{*(1,2)}(N)$, and $c_n^{*(1)}(N)$ obtained for
$n=0,1,2$ in Appendix \ref{sec:numericaldata}. In this table, the data
is given to 6 significant figures for economy of presentation, while
in point of fact, accuracy to 10 significant figures was needed for
the analysis that now follows.  We were able to achieve this numerical
accuracy up to $N=36$ for the $n=0$ mode, and up to $N=26$ for the
modes $n=1$ and $n=2$.  Note that $c_n^{*(0)}(N)$ for $n=0,1,2$
reproduces the results of \cite{ref:ffgw03}.

We now introduce a generalised version of the ansatz first introduced
in \cite{ref:ffgw03}, expanded to O$(1/cL)$ \vspace{1cm}
\begin{eqnarray}
  \label{eq:ansatz}
  c_n(N) &=& A_{\infty,n} \left(1+\frac{\alpha_n N}{c L}\right) N^{\frac{1}{2}+\frac{\beta_n N}{c L}}+C_{\infty,n}\left(1 + \frac{\gamma_n N}{c L}\right)\\ 
  \label{eq:ansatzsplit}&=&  (A_{\infty,n}\sqrt{N} + C_{\infty,n}) + N (A_{\infty,n} \alpha_n \sqrt{N} + A_{\infty,n} \beta_n \sqrt{N} \ln N+ C_{\infty,n}\gamma_n)\left(\frac{1}{cL}\right)+\cinf{2}.
\end{eqnarray}

It is now pertinent to determine bounds on $\alpha_n$ and $\beta_n$,
and a value for $\gamma_n$.  Consider the set of three linear
equations describing the $1/cL$ term of (\ref{eq:ansatzsplit}) for
three consecutive values of $N$

\begin{equation}
  \label{eq:3pointfit}
  \left( \begin{array}{ccc}
      \sqrt{N-1} & \sqrt{N-1} \ln(N-1) & 1\\
      \sqrt{N} & \sqrt{N} \ln(N) & 1\\
      \sqrt{N+1} & \sqrt{N+1} \ln(N+1) & 1
    \end{array}\right)
  \left( \begin{array}{c}
      A_{\infty,n} \alpha_n \\
      A_{\infty,n} \beta_n \\
      C_{\infty,n} \gamma_n
    \end{array} \right) = 
  \left( \begin{array}{c}
      c_n^{*(1)}(N-1)\\
      c_n^{*(1)}(N) \\
      c_n^{*(1)}(N+1)\end{array}\right)
\end{equation}
\\
and the set of two linear equations describing the $1/cL$ term of
(\ref{eq:ansatzsplit}) with $C_{\infty,n}=0$ for two consecutive
values of $N$

\begin{equation}
  \label{eq:2pointfit}
  \left(\begin{array}{cc}
      \sqrt{N} & \sqrt{N}\ln(N) \\
      \sqrt{N+1} & \sqrt{N+1} \ln(N+1)
    \end{array}\right)
  \left(\begin{array}{c}
      A_{\infty,n} \alpha_n\\
      A_{\infty,n} \beta_n
    \end{array}\right) =
  \left(\begin{array}{cc}
      c_n^{*(1)}(N)\\
      c_n^{*(1)}(N+1)
    \end{array}\right).
\end{equation}

Values of $A_{\infty,n}$ and $C_{\infty,n}$ from \cite{ref:ffgw03} are
displayed in Table \ref{tab:ansatzparams}.  The solution for
$\alpha_n$, $\beta_n$, $\gamma_n$, of (\ref{eq:3pointfit}) and the
solution for $\alpha_n$, $\beta_n$ for (\ref{eq:2pointfit}) for
various values of $N$ establish bounds on $\alpha_n$ and $\beta_n$,
and a value for $\gamma_n$.  We establish numerical stability for
these bounds by calculating them for $N=2$ up to $N=35$.  A fractional
accuracy of $\gtrsim 10^{-10}$ is necessary, and hence the parameters
are calculated at $N=35$ for $n=0$, and $N=25$ for $n=1$ and $n=2$.
We present them in Table \ref{tab:ansatzparams}.

\begin{table}[htbp]
  \centering
  \begin{tabular}{cr@{.}lcr@{.}l@{.}lcr@{.}l}
    \hline
    $n$ & \multicolumn{2}{c}{$A_{\infty,n}$} & $C_{\infty,n}$ & \multicolumn{3}{c}{$\alpha_n$} & $\beta_n$ & \multicolumn{2}{c}{$\gamma_n$}\\
    \hline
    \hline\vspace{-0.35cm}\\
    0$\quad$ & 1&54273$\quad$ & $-0.5725\quad$ & 0 &$1561 < \alpha_0 < 0$&$1838\quad$ & $1.998 < \beta_0 < 2.003\quad$ & $0$&$1599$ \vspace{0.06cm}\\
    1$\quad$ & 0&5143$\quad$ & $-0.5739\quad$ & $-5$&$709 < \alpha_1 < -5$&$067\quad$ & $1.972 < \beta_1 < 2.094\quad$ & $-1$&$109$ \vspace{0.06cm}\\
    2$\quad$ & 0&3676$\quad$ & $-0.5775\quad$ & $-8$&$350 < \alpha_2 < -6$&$121\quad$ & $1.887 < \beta_2 < 2.313\quad$ & $-2$&$736$
  \end{tabular}
  \caption{Parameters for the ansatz (\ref{eq:ansatzsplit})}
  \label{tab:ansatzparams}
\end{table}

Note that $\beta_n$ is very close to 2 for $n=0$, and suggestive of
the value 2 for $n=1$ and 2.  We postulate that $\beta_n=2$ for all
$n$. We say more about this $\beta_n$ in Section \ref{sec:conclusion}.

\section{Correlation functions and structure factors}
\label{sec:correlation}

The definition of the two-point correlation function $g_N(x,0)$ is 

\begin{equation}
  \label{eq:correlationdefn}
  g_N(x,0)=\frac{1}{\mathcal N^2} \sum_{j=0}^{N-2}\int_{R_{N-2,j}(x)}\bbs\bbs\bbs dx_1\ldots dx_{N-2} |\psi_N(0,x_1,\ldots,x_j,x,x_{j+1},\ldots,x_{N-2})|^2
\end{equation}
\\
where the normalisation ${\mathcal N^2}$ is specified by
(\ref{eq:normalisation}), and the domain of integration by
(\ref{eq:RNj}).  Then (\ref{eq:correlationdefn}) has the property

\begin{equation}
  \label{eq:gnorm}
  \int_0^L g_N(x,0) dx = N-1,
\end{equation}
\\
in keeping with the interpretation of $g_N(x,0)$ as the density of
particles at position $x$, given there is a particle at the origin.
The structure factor is defined by

\begin{equation}
  \mathcal{S}_n(N) = \frac{1}{N} \frac{1}{\mathcal N^2} \int_{R_{N-1,j}} \bbs\bbs dx_1\ldots dx_{N-1}\left|\sum_{j=0}^{N-1} e^{2\ii\pi  x_j n/L} \right|^2 \left| \psi_N(0,x_1,\ldots,x_{N-1})\right|^2
\end{equation}
\\
where $x_0=0$, which can be written in terms of $g_N(x,0)$ to read

\begin{equation}
  \mathcal{S}_n(N) = 1+\int_0^L g_N(x,0) \cos(2\pi n x/L) dx.
\end{equation}

In view of (\ref{eq:gnorm}) this gives $\mathcal{S}_0(N)=N$.

\subsection{$N=2$}

This subsection yields the most complete set of results within Section
\ref{sec:correlation}, with the set of results for $N=3$ and $N=4$
becoming increasingly limited.  For $N=4$, we display only the small
$cL$ expansion of the correlation function and structure factor.

It is possible to produce a simple, exact form of the correlation
function for $N=2$ by utilising (\ref{eq:BAENis2exact}) to obtain

\begin{equation}
  g_2(x,0)=\frac{2k_2 \cos^2 [\frac{1}{2}k_2 (L-2 x)]}{k_2 L+\sin (k_2 L)}
\end{equation}
\\
which has corresponding structure factor

\begin{equation}
  \label{eq:exactSFNis2moreconcise}
  \mathcal{S}_n(2) = \left\{ \begin{array}{ll}
      2 & \textrm{when } n=0\vspace{0.1cm}\\
      1+\frac{k_2^2 L^2 \sin (k_2 L)}{\left(k_2^2 L^2-n^2 \pi ^2\right) (k_2 L+\sin (k_2 L))}&\textrm{when }n\neq 0.
    \end{array}\right.
\end{equation}

The correlation function, expanded about small $cL$ using
(\ref{eq:BAENis2}), is given by

\begin{eqnarray}
  \label{eq:corrsmallcNis2}
  g_2(x,0) &=& \frac{1}{L}\Big[1+  \left(-\frac{x^2}{L^2}+\frac{x}{L}-\frac{1}{6}\right)(cL)+ \left(\frac{x^4}{3 L^4}-\frac{2 x^3}{3 L^3}+\frac{x^2}{2 L^2}-\frac{x}{6 L}+\frac{1}{60}\right)(cL)^2\nonumber\\
  &&+ \left(-\frac{2 x^6}{45L^6}+\frac{2 x^5}{15L^5}-\frac{7
      x^4}{36L^4}+\frac{x^3}{6L^3}-\frac{7
      x^2}{90L^2}+\frac{x}{60L}-\frac{1}{945}\right)(cL)^3+\czero{4}\Big]
\end{eqnarray}
\\
and hence the structure factor is given by

\begin{equation}
  \mathcal{S}_n(2) = \left\{ \begin{array}{ll}
      2 & \textrm{when } n=0\vspace{0.1cm}\\
      1-\frac{1}{2 n^2 \pi ^2}(cL)+\frac{n^2\pi^2-6}{12n^4\pi^4}(cL)^2-\frac{n^4\pi ^4 -15 n^2\pi ^2 +60}{120 n^6 \pi ^6}(cL)^3+\czero{4} & \textrm{when } n\neq 0.
\end{array} \right. 
\end{equation}

The large $cL$ expansion of the correlation function is given using
(\ref{eq:jimbomiwalargeskappa})

\begin{equation}
  \label{eq:corrNis2largec}
  g_2(t,0)=\frac{1}{L}\left\{ 2\sin^2t + 8 \sin t [ (\pi-2t) \cos t - \sin t]\left(\frac{1}{cL}\right)+\cinf{2}\right\}
\end{equation}
\\
taking the Fourier transform then yields the structure factor

\begin{equation}
  \mathcal{S}_n(2) = \left\{ \begin{array}{ll}
      2 & \textrm{when } n=0\vspace{0.1cm}\\
      \frac{1}{2}+3\left(\frac{1}{c L}\right)+\cinf{2} &\textrm{when } |n|=1\vspace{0.1cm}\\
      1-\frac{4}{(n^2-1)}\left(\frac{1}{cL}\right) +\cinf{2}& \textrm{when } |n|\geq 2.
    \end{array}\right.
\end{equation}

The correlation function for the special values $c L=\pi$, $k_2 L
=\pi/2$, (see Table \ref{tab:BAEexactsolns}) is

\begin{equation}
  g_2(t,0)=\frac{\pi  \left(\sin t +1\right)}{L (\pi
  +2 )}
\end{equation}
\\
and hence the structure factor is

\begin{equation}
  \mathcal{S}_n(2) = \left\{ \begin{array}{ll}
      2 & \textrm{when } n=0\vspace{0.1cm}\\
      1-\frac{2}{(4 n^2-1) (\pi+2 )}& \textrm{when } n\neq 0.
    \end{array} \right.
\end{equation}

\subsection{$N=3$}

Given the length of the full correlation function for $N=3$, it is not
possible to display the full result here, however we shall display the
small and large $cL$ expansions for both the correlation function and
the structure factor.

The small $cL$ expansion of the correlation function is given, using
(\ref{eq:BAENis3})

\begin{eqnarray}
  \label{eq:corrsmallcNis3}
  g_3(x,0) &=& \frac{2}{L}\Big[1+\left(-\frac{ x^2}{L^2}+\frac{ x}{L}-\frac{1}{6}\right)(cL)+ \left(\frac{x^4}{12 L^4}-\frac{x^3}{6 L^3}+\frac{x^2}{4 L^2}-\frac{x}{6L}+\frac{1}{40}\right)(cL)^2\nonumber\\
  && + \left(\frac{x^6}{5L^6}-\frac{3 x^5}{5L^5}+\frac{5
      x^4}{8L^4}-\frac{x^3}{4L^3}+\frac{x}{40L}-\frac{1}{280}\right)(cL)^3+ \czero{4}\Big]
\end{eqnarray} 
\\
and hence structure factor by

\begin{equation}
  \mathcal{S}_n(3) = \left\{ \begin{array}{ll}
      3 & \textrm{when } n=0\vspace{0.1cm}\\
      1-\frac{1}{n^2 \pi ^2}(cL)+\frac{2 n^2 \pi ^2-3}{12 n^4 \pi ^4}(cL)^2-\frac{ n^4\pi ^4 +15 n^2\pi ^2 -180}{40 n^6 \pi ^6}(cL)^3+\czero{4}& \textrm{when } n\neq 0.
\end{array} \right.
\end{equation}

The large $cL$ expansion for the correlation function is given using
(\ref{eq:jimbomiwalargeskappa})

\begin{equation}
  \label{eq:corrNis3largec}
  g_3(t,0)= \frac{2}{L}\left\{\frac{4\left( 2 + \cos2t \right) \sin^2 t}
    {3} - \frac{8\sin t
      \left[ -2\left( \pi  - 2t \right) 
        \left( 2\cos t + \cos3t \right)  +        3\sin t + 4\sin3t \right] }{3}\left(\frac{1}{cL}\right)+\cinf{2}\right\}
\end{equation}
\\
which yields the structure factor

\begin{equation}
  \mathcal{S}_n(3) = \left\{ \begin{array}{ll}
      3 & \textrm{when } n=0\vspace{0.1cm}\\
      \frac{1}{3}+\frac{32}{9 }\left(\frac{1}{cL}\right)+\cinf{2} &\textrm{when } |n|=1\vspace{0.1cm}\\
      \frac{2}{3}+\frac{38}{9}\left(\frac{1}{cL}\right)+\cinf{2} & \textrm{when } |n|=2\vspace{0.1cm}\\
      1-\frac{16 (n^2-2)}{(n^2-4)(n^2-1)}\left(\frac{1}{cL}\right)+\cinf{2} & \textrm{when } |n|\geq 3.
    \end{array}
  \right.
\end{equation}

We now display the correlation function for the special value $k_3
L=\pi$, $c L=\sqrt{2}\pi$, (from Table \ref{tab:BAEexactsolns})

\begin{equation}
  g_3(t,0)= -\frac{2 \pi  \left[2\left(4 \sqrt{2}- \pi \right) \cos t-\pi  \cos 2t-4 \left(2+\sqrt{2} \pi \right) \sin t+2 \left(6+\sqrt{2} \pi \right) \sin 2t-9 \pi -8 \sqrt{2}\right]}{L \left(16+16 \sqrt{2} \pi +9 \pi  ^2\right)}
\end{equation}
\\
and the corresponding structure factor by

\begin{equation}
  \mathcal{S}_n(3) = \left\{ \begin{array}{ll}
      3 & \textrm{when } n=0\vspace{0.1cm}\\
      1-\frac{32+16 \sqrt{2} \pi - 3 \pi ^2}{48+48 \sqrt{2} \pi+27 \pi ^2} & \textrm{when } |n|=1\vspace{0.1cm}\\
      1-\frac{16 \left(2+\sqrt{2} \pi \right)}{\left(4
          n^2-1\right) \left(16+16 \sqrt{2} \pi +9 \pi
          ^2\right)} & \textrm{when }|n|\geq 2.
    \end{array} \right.
\end{equation}

\subsection{$N=4$}

Here we display the correlation function for the small $cL$ expansion
of $N=4$

\begin{eqnarray}
  \label{eq:corrsmallcNis4}
  g_4(x,0) &=& \frac{3}{L}\Big[1+\left(-\frac{ x^2}{L^2}+\frac{ x}{L}-\frac{1}{6}\right)(cL)+ \left(-\frac{x^4}{6 L^4}+\frac{x^3}{3 L^3}-\frac{x}{6
   L}+\frac{1}{30}\right)(cL)^2\nonumber\\
&& + \left(\frac{7 x^6}{18 L^6}-\frac{7 x^5}{6 L^5}+\frac{47
   x^4}{36 L^4}-\frac{2 x^3}{3 L^3}+\frac{19 x^2}{180
   L^2}+\frac{x}{30 L}-\frac{1}{135}\right)(cL)^3+ \czero{4}\Big]
\end{eqnarray} 

and its corresponding structure factor

\begin{equation}
  \mathcal{S}_n(4) = \left\{ \begin{array}{ll}
      4 & \textrm{when } n=0\vspace{0.1cm}\\
      1-\frac{3}{2n^2 \pi ^2}(cL)+\frac{n^2\pi^2+3}{4 n^4 \pi ^4}(cL)^2-\frac{2 n^4\pi ^4 +60 n^2\pi ^2 -525}{40 n^6 \pi ^6}(cL)^3+\czero{4}& \textrm{when } n\neq 0.
    \end{array} \right.
\end{equation}

\subsection{General $N$}

Through close examination of the small $cL$ expansions of the
correlation function for $N=2$ (\ref{eq:corrsmallcNis2}), $N=3$
(\ref{eq:corrsmallcNis3}) and $N=4$ (\ref{eq:corrsmallcNis4}), we find
the following polynomial structure

\begin{eqnarray}
\label{eq:correlationsmallc}
  g_N(x,0) &=& \frac{N-1}{L} \Big\{1 -  f(x)(cL) +  \left[ \left(-\frac{N}{4}+\frac{5}{6}\right) f(x)^2+\left(\frac{N}{12}-\frac{1}{9}\right) f(x) + \left(\frac{N}{720}-\frac{1}{216}\right)\right](cL)^2 \nonumber\\ 
  &&+\Big[\left(-\frac{N^2}{36}+\frac{23N}{60} -\frac{7}{10}\right)f(x)^3 + \left(\frac{N^2}{36} - \frac{7N}{40}+\frac{1}{5}\right)f(x)^2 \nonumber\\
  &&+ \left(-\frac{N^2}{144}+\frac{13N}{720}-\frac{1}{120}\right)f(x)+\left(-\frac{N^2}{6804}+\frac{79N}{90720}-\frac{1}{1080}\right)\Big](cL)^3+\czero{4}\Big\}
\end{eqnarray}
\\
where $f(x)=x^2/L^2 - x/L + 1/6$.  We conjecture that this structure
continues for all $N$.  This has corresponding structure factor

\begin{equation}
  \mathcal{S}_n(N) = \left\{\begin{array}{ll}
      N & \textrm{when } n=0\vspace{0.1cm}\\
      1-\frac{(N-1)}{2n^2\pi^2}(cL)+\frac{(N-1) \left(2 n^2\pi^2+9
          N-30\right)}{24 n^4 \pi ^4}(cL)^2&\vspace{0.1cm}\\
      \quad\quad-\frac{(N-1) \left(N n^4\pi ^4 +75
          N n^2\pi^2-180 n^2\pi^2+75
          N^2-1035 N+1890\right)}{240 n^6 \pi
        ^6}(cL)^3+\czero{4} &\textrm{when } n\neq 0
\end{array} \right.
\end{equation}

We have not proceeded further than $N=3$ for the large $cL$ expansion
for the same lack of utility as we encountered in the case of the
density matrix in Section \ref{sec:density}.  There is now,
once again, a powerful way to proceed to develop such an expansion for
general $N$, and we present the derivation in the following section.
The special case of $N=2$ (\ref{eq:corrNis2largec}) and $N=3$
(\ref{eq:corrNis3largec}) provide useful checks.

In closing this section, we note that in (\ref{eq:correlationsmallc}),
the leading order term for $g_N(0,0)$ is $(N-1)/L$, which is the free
Bose result.  Deviations from this begin at order $cL$.

\section{Correlation functions and structure factors for large $cL$}
\label{sec:correlationlargec}

In Section \ref{sec:toeplitz} results from Jimbo and Miwa
\cite{ref:jimbomiwa} were used to express the O$(1/cL)$ correction to
the density matrix in the form of Toeplitz determinants, which could
then be numerically analysed.  The method of \cite{ref:jimbomiwa} can
also be applied to the calculation of the O$(1/cL)$ correction to the
two-point correlation function (\ref{eq:correlationdefn}), yielding in
fact a closed form analytic expression.  To derive this, we follow the
working in \cite{ref:jimbomiwa}, and begin by noting that the
O$(1/cL)$ expansion of (\ref{eq:wavefunctiondefn}) is

\begin{equation}
  \label{eq:wavefunctionlargec}
  \psi_N(x_1,x_2,\ldots,x_N) = \left(1+\frac{1}{cL}\left\{ \sum_{l=1}^N (-N+2l-1) \dd{x_l} - 2\rho_0 \sum_{l=1}^N x_l \dd{x_l}\right\}+\ldots\right) \det[e^{\ii k_j x_l}]
\end{equation}

Now put $N \to N+2$ (for convenience) and label the particles as

\begin{equation}
  \label{eq:np2order}
  0 < y < x_1 < \ldots < x_j < x < x_{j+1} < \ldots < x_N < L.
\end{equation}

In the definition of $g_{N+2}(x,0)$, $x$ will be the variable as in
(\ref{eq:np2order}) and we will take $y\to 0$.  With these labels and
$N \to N+2$ the operator in (\ref{eq:wavefunctionlargec}) reads

\begin{equation}
  \label{eq:newoperator}
  -(N+1) \dd{y} + \sum_{l=1}^j(-N-1+2l)\dd{x_l}+(-N+1+2j) \dd{x} + \sum_{l=j+1}^N(-N+1+2l) \dd{x_l} - 2\rho_0 \left(\sum_{l=1}^N x_l \dd{x_l}+x \dd{x} + y \dd{y}\right).
\end{equation}

The determinant in (\ref{eq:wavefunctionlargec}) has the translation
invariance property

\begin{equation}
  \det\left[\begin{array}{l}
      e^{\ii k_j y} \\
      e^{\ii k_j x_k}
    \end{array} \right] = 
  \det\left[\begin{array}{l}
      1\\
      e^{\ii k_j (x_k-y)}
    \end{array} \right]
\end{equation}
\\
since $\sum k_j =0 $.  This means that we can write

\begin{equation}
  \dd{y} = - \dd{x}-\sum_{j=1}^N \dd{x_j}
\end{equation}
\\
and using too the fact that we want $y\to 0$, the operator (\ref{eq:newoperator}) reads

\begin{equation}
  \sum_{l=1}^j 2l \dd{x_l}+2 (1+j) \dd{x} + \sum_{l=j+1}^N 2 (1+l) \dd{x_l} - 2\rho_0 \left(\sum_{l=1}^N x_l \dd{x_l}+x \dd{x} \right).
\end{equation}

In keeping with (\ref{eq:correlationdefn}), by definition

\begin{equation}
  \label{eq:correlationnewdef}
  \pairatc = \frac{1}{{\mathcal N}^2} \lim_{y\to 0} \sum_{j=0}^N\int_{R_{N,j}(y,x)} \bbs\bbs\bbs dx_1 \ldots dx_N \left| \psi_{N+2} (y,x_1,\ldots,x_j,x,x_{j+1},\ldots,x_N)\right|^2
\end{equation}
\\
where ${\mathcal N}^2$ is the normalisation, defined by
(\ref{eq:normalisation}), and $R_{N,j}(y,x)$ is the region of
integration specified by

\begin{equation}
  R_{N,j}(y,x):0\leq y \leq x_1 \leq \ldots  \leq x_j \leq x \leq x_{j+1} \leq \ldots \leq x_N \leq L.
\end{equation}

With

\begin{equation}
  (\tilde{x}_1,\tilde{x}_2,\ldots,\tilde{x}_{N+2}) = (0,x_1,\ldots,x_j,x,x_{j+1},\ldots,x_N)
\end{equation}
\\
and

\begin{eqnarray}
  A &=& \sum_{l=1}^N x_l \dd{x_l} + x \dd{x}\\
  B_j &=& \sum_{l=1}^j 2 l \dd{x_l} + 2(1+j) \dd{x} + \sum_{l=j+1}^N 2(1+l) \dd{x_l}
\end{eqnarray}
\\
we have

\begin{eqnarray}
  \label{eq:limeq}
  \lim_{y\to 0} |\psi_{N+2}(y,x_1,\ldots,x_j,x,x_{j+1},\ldots,x_N)|^2 &=& \detplanesq - \frac{2 \rho_0}{c } A \left( \detplanesq\right)\nonumber\\
  &&+\frac{1}{c} B_j \left(\detplanesq\right) + \cinf{2}.
\end{eqnarray}

The normalisation (\ref{eq:normalisation}) was first expanded in large
$cL$ by \cite{ref:jimbomiwa}, we display here the first two orders

\begin{equation}
  \label{eq:normlargec}
  {\mathcal N}^2 = ({\mathcal N^{(\infty)}})^2 \left[ 1+\frac{2 \rho_0}{c} (N+1) + \cinf{2}\right].
\end{equation}
\\
where $({\mathcal N^{(\infty)}})^2$ is the normalisation
(\ref{eq:normalisation}) in the limit $cL\to\infty$.  Substituting
(\ref{eq:limeq}) and (\ref{eq:normlargec}) in
(\ref{eq:correlationnewdef}) shows that to O$(1/cL)$

\begin{eqnarray}
  \pairatc &=&  \pairatinf -\frac{2\rho_0}{c}(N+1) \pairatinf\\
  && -\frac{2\rho_0}{c} \ninfinv \sum_{j=0}^N\int_{R_{N,j}(x)}\bbs\bbs dx_1 \ldots dx_N A \left(\detplanesq\right)\\
  && + \frac{1}{c}\ninfinv\sum_{j=0}^N \int_{R_{N,j}(x)} \bbs\bbs dx_1 \ldots dx_N B_j \left(\detplanesq\right).
\end{eqnarray}

To proceed further consider $\detplanesq$ as a function of $x_l$,
$(l=1,\ldots,N)$.  This function vanishes at the three points
$x_l=0,x_j,L \,\;(j\neq l)$.

It follows that

\begin{equation}
  \int_{R_{N,j}(x)}\bbs\bbs dx_l\; x_l \dd{x_l}\detplanesq = - \int_{R_{N,j}(x)}\bbs\bbs dx_l \detplanesq
\end{equation}
\\
while

\begin{equation}
  \int_{R_{N,j}(x)} \!\bbs dx_l \dd{x_l}\detplanesq = 0.
\end{equation}

Hence

\begin{eqnarray}
  \pairatc &=&\pairatinf -\frac{2\rho_0}{c} (N+1) \pairatinf -\frac{2\rho_0}{c} \left( x \dd{x} - N\right) \pairatinf\\
  &&+\frac{1}{c}\ninfinv \sum_{j=0}^N 2(j+1) \dd{x}\int_{R_{N,j}(x)}\bbs\bbs dx_1\ldots dx_N \detplanesq\\
  &=& \pairatinf-\frac{2\rho_0}{c} \left( x \dd{x} +1\right) \pairatinf\label{eq:pairatinffirstline}\\
  && + \frac{2}{c} \ninfinv \dd{x} \sum_{j=0}^N \frac{(j+1)}{j!(N-j)!}\int_{R_{N,j}(x)} \bbs\bbs dx_1 \ldots dx_N \detplanesq\label{eq:pairatinflastline}
\end{eqnarray}
\\
where (\ref{eq:pairatinflastline}) can be written

\newcommand{\frontfac}{\frac{2}{c}\ninfinv \frac{1}{N!}}

\begin{align}
& \frontfac \dd{x}\dd{\xi}\sum_{j=0}^N \xi^{j+1} \left(\begin{array}{l}N\\j\end{array}\right) \left.\manyints \detplanesq\right|_{\xi=1}\\
 & \quad=\frontfac \dd{x}\dd{\xi}\xi \prod_{l=1}^N \left.\left( \int_x^L + \xi \int_0^x\right) dx_l \detplanesq\right|_{\xi=1}\\
 & \quad=\frontfac \dd{x} \int_0^L dx_1\ldots dx_N \detplanesq \label{eq:firstline}\\
 &\quad\quad +\frontfac\dd{x}\dd{\xi}\prod_{l=1}^N \left.\left(\int_x^L+\xi \int_0^x\right) dx_l \detplanesq\right|_{\xi=1}\label{eq:secondline}
\end{align}
\\
we observe that

\begin{equation}
  \label{eq:pairatinfnice}
  \frac{1}{N!}\ninfinv \int_0^L dx_1 \ldots dx_N \detplanesq = \pairatinf = \rho_0 (1-y(x)^2)
\end{equation}
\\
where

\begin{equation}
  \label{eq:y(x)}
  y(x) = \frac{\sin\left(\frac{(N+2) \pi x}{L}\right)}{(N+2) \sin\left(\frac{\pi x}{L}\right)} = \frac{1}{\rho_0} K_{N+2,L}(x,0)
\end{equation}
\\
where we note that (\ref{eq:pairatinfnice}) and (\ref{eq:y(x)}) also
appear in the CUE (Circular Unitary Ensemble), given by Dyson
\cite{ref:dyson}.

Hence (\ref{eq:firstline}) is equal to

\begin{equation}
  \frac{2}{c} \dd{x} \pairatinf.
\end{equation}

Regarding (\ref{eq:secondline}), note that

\begin{eqnarray}
  &&  \dd{\xi} \prod_{l=1}^N \left.\left(\int_x^L+\xi\int_0^x\right)dx_l \detplanesq\right|_{\xi=1}\\
  &&\quad=\dd{\xi}\left.\left(\prod_{l=1}^N\int_0^L dx_l \detplanesq + \sum_{l=1}^N \int_0^x dx_l (\xi-1) \prod_{j=1,j\neq l}^N \int_0^L dx_j \detplanesq + O((\xi-1)^2)\right)\right|_{\xi=1}\\
  &&\quad= \sum_{l=1}^N \int_0^x dx_l \prod_{j=1,j\neq l}^N \int_0^L dx_j \detplanesq\\
  &&\quad= N \int_0^x dx_1 \int_0^L dx_2 \ldots \int_0^L dx_N \detplanesq.
\end{eqnarray}

Therefore, (\ref{eq:secondline}) is equal to 

\begin{equation}
  \label{eq:pairatinflastlineequiv}
  \frac{2}{c}\ninfinv \frac{1}{(N-1)!} \dd{x} \int_0^x dx_1 \int_0^L dx_2 \ldots \int_0^L dx_N \detplanesq.
\end{equation}

But for the free Fermi system the 3-point correlation function is
specified by

\begin{eqnarray}
  g_{3,(N+2)}^{(\infty)}(x,0,x_1) &=& \frac{1}{(N-1)!}\ninfinv \int_0^L dx_2 \ldots \int_0^L dx_N \detplanesq\\
  &=& \rho_0^2\det\left(\begin{array}{ccc}
      1     & y(x)    & y(x_1)\\
      y(x)  & 1       & y(x-x_1)\\
      y(x_1)& y(x-x_1)& 1\\
    \end{array}\right)\\
  &=& \rho_0^2[1- y(x)^2-y(x_1)^2-y(x-x_1)^2+2y(x)y(x_1)y(x-x_1)]\label{eq:g3spread}
\end{eqnarray}
\\
where $y(x)$ is specified by (\ref{eq:y(x)}). Note that this
definition is only valid for $N\geq 1$, a 3-point correlation function
for a system of 2 particles does not make physical sense,
$g^{(\infty)}_{3,2}(x,0,x_1)=0$.  Now
(\ref{eq:pairatinflastlineequiv}) reduces to

\begin{equation}
  \frac{2}{c} \dd{x} \int_0^x dx_1 g_{3,(N+2)}^{(\infty)}(x,0,x_1).
\end{equation}

Finally, adding up all contributions gives the sought closed form
expression (here we revert to $N+2\to N$)

\begin{equation}
  g_{N}(x,0) = g_{N}^{(\infty)}(x,0) - \frac{2\rho_0}{c}\left(x \dd{x} + 1\right)g_{N}^{(\infty)}(x,0)+\frac{2}{c}\dd{x} \pairatinfatN + \frac{2}{c}\dd{x}\int_0^x dx_1 g_{3,N}^{(\infty)}(x,0,x_1)
\end{equation}
\\
correct to O$(1/cL)$, valid for all $N\geq2$. This has been checked
with (\ref{eq:corrNis2largec}) and (\ref{eq:corrNis3largec}).  Using
(\ref{eq:pairatinfnice}) and (\ref{eq:g3spread}) this can be written
in the simpler form

\begin{eqnarray}
  \label{eq:ginys}
  g_N(x,0) &=& \pairatinfatN + 4 N \left(\frac{1}{cL}\right) \Big\{- y(x)\left[\rho_0 y(x)+ y'(x)\right]\nonumber\\
  &&+\rho_0 \dd{x}\left[y(x) \int_0^x  y(x_1) y(x-x_1) dx_1\right]\Big\}+\cinf{2}
\end{eqnarray}
\\
which is valid for all $N\geq3$.  We have confirmed that this result
recovers (\ref{eq:corrNis3largec}).  Now (\ref{eq:ginys}), and
concomitantly its structure factor, are readily computed for any
$N\geq3$.  The structure factor in the limit $cL\to\infty$, is given
by

\begin{equation}
  \mathcal{S}_n(N) = \left\{\begin{array}{ll}
      N & \textrm{when } n=0\\
      |n|/N & \textrm{when } 0 < |n| < N\\
      1 & \textrm{when } |n| \geq N.
\end{array}\right.
\end{equation}

In the thermodynamic limit, $y(x)$ becomes

\begin{equation}
  \label{eq:y(x)TDL}
  y(x) = \frac{\sin(\rho_0\pi x)}{\rho_0\pi x}
\end{equation}
\\
and now the correlation function follows from (\ref{eq:ginys}) and is
(where we use the appropriate scaled variable $\rho_0/c$)

\begin{eqnarray}
  \label{eq:correlationTDL}
  g_{\infty}(x,0)&=& \rho_0\left(1 -\frac{\sin ^2\bar{x}}{\bar{x}^2}\right)-4\rho_0 \Big\{ \frac{\sin \bar{x}}{\bar{x}^3} \left[\pi  \bar{x}\cos \bar{x}+(\bar{x}-\pi ) \sin\bar{x}\right]\nonumber\\
  &&-\dd{x}\left[y(x)\int_0^x y(x_1) y(x-x_1) dx_1\right]\Big\}\left(\frac{\rho_0}{c}\right) + \textrm{O}\left(\frac{\rho_0}{c}\right)^2
\end{eqnarray}
\\
where $\bar{x}=\rho_0\pi x$.  This equation was first given by Korepin
\cite{ref:korepin}.  Recently in \cite{ref:chernybrand}, using the Random
Phase Approximation (RPA), which they indicated is valid to the 
$\rho_0/c$ correction,  the structure factor was calculated in the
thermodynamic limit from which (\ref{eq:correlationTDL}) is recovered.

Here we explicitly evaluate the integral in (\ref{eq:correlationTDL})
with (\ref{eq:y(x)TDL})
 
\begin{equation}
  \int_0^x y(x_1) y(x-x_1) dx_1=\frac{\sin( \bar{x}) \text{Si}(2 \bar{x} )+\cos(\bar{x}) [\text{Ci}(2 \bar{x})-\log (2\bar{x})-\gamma ]}{\rho_0\pi   \bar{x}}
\end{equation}
\\
where $\gamma$ is Euler's constant, and Si and Ci are the sine
integral and cosine integral functions, respectively.

In closing this section we note that $g_N(0,0)$ in (\ref{eq:ginys})
and similarly (\ref{eq:correlationTDL}) is zero to leading order, the
free Fermi result.  Deviations from zero do not begin until higher
order than $1/cL$ as can be seen from these equations.

\section{Concluding remarks}
\label{sec:conclusion}

The occupation numbers for the large $cL$ limit given in
(\ref{eq:ansatz}) can be continued to the asymptotically large $N$
limit giving

\begin{equation}
  \label{eq:occNumLargeN}
  c_n(N) \sim N^{\frac{1}{2}+\frac{\beta_n N}{cL}}.
\end{equation}
\\
Very strong evidence for the exponent $\beta_n$ having the integer
value 2 was presented.

In constructing the ansatz for the occupation numbers, we chose to
scale $cL$ by $N$ from the outset. This was done because the $cL$
large limit is now compatible with the thermodynamic limit (unlike the
small $cL$ limit as discussed after (\ref{eq:generalc_0(N)})) and
further to this we found that the numerical analysis, for any finite
$N$, in constructing the ansatz, failed without this scaling.

The question then is, how does (\ref{eq:occNumLargeN}) compare with
the thermodynamic limit. In a very nice work, at a time coincident
with the seminal work of the Japanese group
\cite{ref:jimbomiwamorisato,ref:jimbomiwa}, the density matrix first
given by Lenard in the impenetrable limit
\addtocounter{footnote}{-1}\cite{ref:lenard}, \footnotemark{}, was
extended to the $1/c$ correction \cite{ref:creamerthackerwilkinson}.
They used the quantum inverse-scattering method in concert with the
very important work \cite{ref:jimbomiwamorisato} on the impenetrable
Bose gas in terms of Painlev\'{e} V theory, and obtained

\begin{equation}
  \label{eq:densitypain}
  \rho(\mathbf{x}) \sim \frac{1}{|\mathbf{x}|^{\frac{1}{2}-\frac{2k_{\rm F}}{\pi c}}}.
\end{equation}

The Fourier transform of (\ref{eq:densitypain}) gives the momentum
distribution, with the Fermi momentum, $k_{\rm F} = \pi\rho_0$,

\begin{equation}
\label{eq:FTpain}
c(\mathbf{k}) \sim \frac{1}{|\mathbf{k}|^{\frac{1}{2}+\frac{2k_{\rm F}}{\pi c}}}.
\end{equation}

Following the presentation given in \cite{ref:ffgw03}, we can readily
see that (\ref{eq:occNumLargeN}) is one-to-one with (\ref{eq:FTpain}).
An elementary way to see this immediately is to observe that
$|\mathbf{k}|\sim 1/L$ and thus in the thermodynamic limit
$|\mathbf{k}|\sim 1/N$.  Note the integer value of 2 for the
coefficient of $k_{\rm F}/\pi c$ in the exponent
(\ref{eq:densitypain}) and (\ref{eq:FTpain}).

In our work on the impenetrable Bose gas \cite{ref:ffgw03,ref:ffg}, we
studied the system in all boundary conditions, periodic, Dirichlet,
Neumann, as well as for the harmonically trapped system. In all cases,
we found that in the asymptotically large $N$ limit that
(\ref{eq:occNumLargeN}) held in the limit $c\to\infty$.  This firmly
suggests that the exponent obtained, in this large $N$ limit, is
universal, being the same for all boundary conditions and (low lying)
$n$ modes.  Therefore, we anticipate the same is true now for our
(\ref{eq:occNumLargeN}).

The use of periodic boundary conditions, while facilitating the
mathematics, is, nonetheless, a powerful preemptor of the analytical
properties of the Bose gas, a system of both continuing and immense
attraction to the theoretician and the experimentalist, in concert.

\bibliographystyle{apsrev}

\begin{thebibliography}{99}
\bibitem{ref:girardeau} M. Girardeau, \JMP {\bf 1}, 516 (1960).
\bibitem{ref:liebliniger} E. H. Lieb and W. Liniger, \PR {\bf 130}, 1605 (1963).
\bibitem{ref:liebII} E. H. Lieb, \PR {\bf 130}, 1616 (1963).
\bibitem{ref:mugasnider} J. G. Muga and R. F. Snider, \PRA {\bf 57}, 3317 (1998).
\bibitem{ref:sakmann} K. Sakmann, A. I. Streltsov, O. E. Alon, and L. S. Cederbaum, \PRA {\bf 72}, 033613 (2005).
\bibitem{ref:batchelorhard} M. T. Batchelor, X. -W. Guan, N. Oelkers, and C. Lee, \JPA {\bf 38}, 7787 (2005).
\bibitem{ref:lenard} A. Lenard, \JMP {\bf 5}, 930 (1964).
\bibitem{ref:jimbomiwa} M. Jimbo and T. Miwa, \PRD {\bf 24}, 3169 (1981).
\bibitem{ref:korepinbook} V. Korepin, N. Bogoliubov, and A. Izergin, {\it Quantum inverse scattering method and correlation functions} (Cambridge University Press, 1993).
\bibitem{ref:ffgw03} P. J. Forrester, N. E. Frankel, T. M. Garoni, and N. S. Witte, \PRA {\bf 67}, 043607 (2003).
\bibitem{ref:ffg} P. J. Forrester, N. E. Frankel and T. M. Garoni, \JMP {\bf 44}, 4157 (2003).
\bibitem{ref:ffgw03pain} P. J. Forrester, N. E. Frankel, T. M. Garoni and N. S. Witte, \CMP {\bf 238}, 257 (2003).
\bibitem{ref:olshanii} M. Olshanii, \PRL {\bf 81}, 938 (1998).
\bibitem{ref:stoferleretal} T. St\"{o}ferle, H. Moritz, C. Schori, M. K\"{o}hl, and T. Esslinger, \PRL {\bf 92}, 130403 (2004).
\bibitem{ref:tolraetal} B. L. Tolra, K. M. O'Hara, J. H. Huckans, W. D. Phillips, S. L. Rolston, and J. V. Porto, \PRL {\bf 92}, 190401 (2004).
\bibitem{ref:paredesetal} B. Paredes, A. Widera, V. Murg, O. Mandel, S. F\"{o}lling, I. Cirac, G. Shlyapnikov, T. H\"{a}nsch, and I. Bloch, Nature {\bf 429}, 277 (2004).
\bibitem{ref:kinoshitawengerweiss} T. Kinoshita, T. Wenger, and D. Weiss, Science {\bf 305}, 1125 (2004).
\bibitem{ref:kohletal} M. K\"{o}hl, T. St\"{o}ferle, H. Moritz, C. Schori, and T. Esslinger, App. Phys. B {\bf 79}, 1009 (2004).
\bibitem{ref:fertigetal} C. D. Fertig, K. M. O'Hara, J. H. Huckans, S. L. Rolston, W. D. Phillips, and J. V. Porto, \PRL {\bf 94}, 120403 (2005).
\bibitem{ref:bethe} H. Bethe, Z. Physik {\bf 71}, 205 (1931).
\bibitem{ref:baxter} R. J. Baxter, \CJP {\bf 30}, 1005 (1992).
\bibitem{ref:habandhisphysics} C. N. Yang, in {\it Hans Bethe and His Physics}, edited by G. E. Brown and C.-H. Lee (World Scientific Publishing, 2006).
\bibitem{ref:gaudin} M. Gaudin, \PRA {\bf 4}, 386 (1971).
\bibitem{ref:szego} G. Szeg\"{o}, {\it Orthogonal Polynomials} (American Mathematical Society, Providence, RI, 1975).
\bibitem{ref:forresterbook} P. J. Forrester, {\it Log-gases and Random Matrices}, www.ms.unimelb.edu.au/\~{}matpjf.matpjf.html.
\bibitem{ref:dyson} F. J. Dyson, \JMP {\bf 3}, 140, 157, 166 (1962).
\bibitem{ref:korepin} V. E. Korepin, \CMP {\bf 94}, 93 (1984).
\bibitem{ref:chernybrand} A. Y. Cherny and J. Brand, \PRA {\bf 73}, 023612 (2006).
\bibitem{ref:jimbomiwamorisato} M. Jimbo, T. Miwa, Y. Mori and M. Sato, \PhysicaD {\bf 1}, 80 (1980).
\bibitem{ref:creamerthackerwilkinson} D. B. Creamer, H. B. Thacker, and D. Wilkinson, \PRD {\bf 23}, 3081 (1981).
\end{thebibliography}

\newcommand{\JMP}{J. Math. Phys. }
\newcommand{\PRL}{Phys. Rev. Lett. }
\newcommand{\PR}{Phys. Rev. }
\newcommand{\PRA}{Phys. Rev. A }
\newcommand{\PRD}{Phys. Rev. D }
\newcommand{\CMP}{Commun. Math. Phys. }
\newcommand{\PhysicaA}{Physica A }
\newcommand{\PhysicaD}{Physica D }
\newcommand{\CJP}{Chin. J. Phys. }
\newcommand{\JPA}{J. Phys. A }

\appendix
\section{Bethe equation solutions}
\label{sec:smallc}

$N=2$

\begin{eqnarray}
\label{eq:BAENis2}
k_2 = \sqrt{\frac{c}{L}} \left[1- \frac{1}{24}(cL)+\frac{11}{5760}(cL)^2- \frac{17}{322560}(cL)^3 -\frac{281}{154828800}(cL)^4+\czero{5}\right]
\end{eqnarray}

$N=3$

\begin{equation}
\label{eq:BAENis3}
k_3=\sqrt{\frac{3c}{L}} \left[1-\frac{1}{24}(cL)+\frac{19}{5760}(cL)^2-\frac{299}{967680}(cL)^3+\frac{11077}{464486400}(cL)^4+\czero{5}\right]
\end{equation}

$N=4$

\begin{subequations}
\label{eq:BAENis4}
\begin{eqnarray}
  k_4 &=& \sqrt{\left(3+\sqrt{6}\right) \frac{c}{L}} \Big[1-\frac{1}{24}(cL)+\frac{31-2\sqrt{6}}{5760}(cL)^2\\
&&+\frac{-879+86 \sqrt{6}}{967680}(cL)^3+\frac{63381-5500 \sqrt{6}}{464486400}(cL)^4+\czero{5}\Big]\\
  k_3 &=& \sqrt{\left(3-\sqrt{6}\right) \frac{c}{L}}  \Big[1-\frac{1}{24}(cL)+\frac{31+2 \sqrt{6}}{5760}(cL)^2\\
&&+\frac{-879-86 \sqrt{6}}{967680}(cL)^3+\frac{63381+5500 \sqrt{6}}{464486400}(cL)^4+\czero{5}\Big]
\end{eqnarray}
\end{subequations}

$N=5$

\begin{subequations}
\label{eq:BAENis5}
\begin{eqnarray}
  k_5 &=& \sqrt{\left(5+\sqrt{10}\right) \frac{c}{L}} \Big[ 1-\frac{1}{24}(cL)+\frac{39-2\sqrt{10}}{5760}(cL)^2\\
&&+\frac{-1511+118\sqrt{10}}{967680}(cL)^3+\frac{165589 - 13196\sqrt{10}}{464486400}(cL)^4+\czero{5}\Big]\\
  k_4 &=& \sqrt{\left(5-\sqrt{10}\right) \frac{c}{L}} \Big[ 1-\frac{1}{24}(cL)+\frac{39 + 2\sqrt{10} }{5760}(cL)^2\\
&&+\frac{-1511 - 118\sqrt{10}}{967680}(cL)^3+\frac{165589 + 13196\sqrt{10}}{464486400}(cL)^4+\czero{5}\Big]
\end{eqnarray}\end{subequations}

We have also computed these expansions out to $N=10$, due to the
inordinate complexities of the numbers, the expansions for $N\geq6$
are known only in decimal form.

The leading term of $k_j$ is precisely related to the $j$th zero of
the $N$th polynomial, as mentioned in Section
\ref{sec:constructingthewavefunction}.  Note also the universality of
the coefficient of the $(cL)$ term, that is $-1/24$ (as in
(\ref{eq:smallck})).

\section{Fredholm determinants}
\label{sec:fredholm}

Here the Fredholm minor (\ref{eq:fredonebigdef}) will be related to a
multiple integral, which in turn implies the Toeplitz determinant form
(\ref{eq:fredonedef}).  Consider the multiple integral

\begin{eqnarray}
\label{eq:Amultipleintegral}
A_N(x,y) &=& \left(\int_0^L + \lambda \int_y^x \right) dx_2 \ldots \left( \int_0^L + \lambda \int_y^x\right) dx_N \nonumber\\
&&\times \prod_{j=2}^N 2 \sin[(x-x_j)/L] 2 \sin[\pi(y-x_j)/L] \prod_{2\leq j < k \leq N} \left\{ 2 \sin[\pi(x_k-x_j)/L]\right\}^2
\end{eqnarray}

Setting

\begin{equation}
  g(u) = (1+\lambda \chi^{(u)}_{\left[y,x\right]}) 2 \sin\left[\pi(x-u)/L\right] 2 \sin\left[\pi(u-y)/L\right]
\end{equation}
\\
where $\chi^{(u)}_{\left[y,x\right]}=1$ for $u\in[y,x]$ and 0
otherwise, allows this to be written

\begin{equation}
  A_N(x,y) = \int_0^L dx_2 \ldots \int_0^L dx_N \prod_{l=2}^N g(x_l) \prod_{1\leq j < k \leq N} |e^{2\ii\pi  x_k/L}-e^{2\ii\pi x_j/L}|^2.
\end{equation}

By a well known identity (see e.g.~Szeg\"{o} \cite{ref:szego}) this is equal to a Toeplitz determinant, 

\begin{equation}
  \label{eq:Antoeplitzform}
  A_N(x,y) = (N-1)! \det\left[ \int_0^L du \, g(u) e^{2\ii\pi u(j-k)/L}\right]_{j,k=1,\ldots,N-1}.
\end{equation}

On the other hand the integral in $A_N(x,y)$ can be expanded as a
power series in $\lambda$. For this define

\begin{equation}
  \label{eq:phidef}
  \phi_N(x_1,x_2,\ldots,x_N) = \prod_{1\leq j < k \leq N} (e^{2\ii\pi x_k/L} - e^{2\ii\pi x_j/L})
\end{equation}
\\
and introduce the free Fermi type distribution

\begin{eqnarray}
  \label{eq:rhoffdef}
  \rho_N^{\textrm{FF}}(x,y;x_2,\ldots,x_n) &=& \frac{(N-1)!}{(N-n)!C_{N,L}} \int_0^L dx_{n+1} \ldots \int_0^L dx_N\nonumber\\
  &&\times \phi_N(x,x_2,\ldots,x_N)\overline{\phi(y,x_2,\ldots,x_N)}
\end{eqnarray}
\\
where $C_{N,L}$ is such that

\begin{equation}
  \rho_N^{\textrm{FF}}(x,x) = \frac{N}{L}
\end{equation}
\\
and thus given by

\begin{equation}
  \label{eq:ncdef}
  N C_{N,L} = \int_0^L dx_1 \ldots \int_0^L dx_N |\phi(x_1,\ldots,x_N)|^2.
\end{equation}

Now, writing the integrand in (\ref{eq:Amultipleintegral}) in terms of
(\ref{eq:phidef}), expanding in $\lambda$, and making use of the
definition (\ref{eq:rhoffdef}) we see

\begin{equation}
  \label{eq:Andifferent}
  A_N(x,y) = C_{N,L} \sum_{n=0}^{\infty} \frac{\lambda^n}{n!} \int_y^x dx_2 \ldots \int_y^x dx_{n+1} \rho_N^{\textrm{FF}}(x,y; x_2,\ldots,x_{n+1}).
\end{equation}

A straightforward calculation (see e.g.~Forrester
\cite{ref:forresterbook}) gives the determinant form

\begin{equation}
  \label{eq:rhoffmatrix}
  \rho_N^{\textrm{FF}}(x,y;x_2,\ldots,x_{n+1}) = \det\left[\begin{array}{ll}
      K_{NL}(x,y) & \left[K_{NL}(x,u_k)\right]_{k=1,\ldots,n}\\
      \left[K_{NL}(u_j,y)\right]_{j=1,\ldots,n}& \left[K_{NL}(u_j,u_k)\right]_{j,k=1,\ldots,n}
    \end{array}
  \right]
\end{equation}
\\
and furthermore shows from (\ref{eq:ncdef}) that

\begin{equation}
  \label{eq:csimple}
  C_{N,L} = (N-1)! L^N.
\end{equation}

Substituting (\ref{eq:csimple}) and (\ref{eq:rhoffmatrix}) in
(\ref{eq:Andifferent}) and comparing with (\ref{eq:fredonebigdef})
shows

\begin{equation}
  \frac{1}{\lambda} \fredone = \frac{1}{(N-1)!L^N} A_N(x,0).
\end{equation}

With $A_N(x,y)$ in its Toeplitz determinant form
(\ref{eq:Antoeplitzform}), this gives (\ref{eq:fredonedef}).

\newpage
\section{Numerical data}
\label{sec:numericaldata}

\begin{table}[htbp]
  \centering
\begin{sideways}
  \begin{tabular}{|c|cccc|cccc|cccc|}
    \hline
    \hline
    $N$ & $c_0^{*(0)}(N)$ & $c_0^{*(1,1)}(N)$ & $c_0^{*(1,2)}(N)$ & $c_0^{*(1)}(N)$& $c_1^{*(0)}(N)$ & $c_1^{*(1,1)}(N)$ & $c_1^{*(1,2)}(N)$ & $c_1^{*(1)}(N)$& $c_2^{*(0)}(N)$ & $c_2^{*(1,1)}(N)$ & $c_2^{*(1,2)}(N)$ & $c_2^{*(1)}(N)$\\
    \hline
    2   & $0.810569\sldots$   & $-0.757722\sldots$   & $4$   & $3.24228\sldots$   & $0.0900633\sldots$   & $-0.227763\sldots$   & $-1.33333\sldots$   & $-1.56110\sldots$   & $0.00360253\sldots$   & $0.219594\sldots$   & $-0.266667\sldots$   & $-0.0470731\sldots$   \\
    3   & $0.702151\sldots$   & $-1.78709\sldots$   & $8$   & $6.21291\sldots$   & $0.111111\sldots$   & $-0.924374\sldots$   & $-1$   & $-1.92437\sldots$   & $0.0328356\sldots$   & $0.249760\sldots$   & $-1.33333\sldots$   & $-1.08357\sldots$   \\
    4   & $0.629414\sldots$   & $-2.96469\sldots$   & $11.9525\sldots$   & $8.98781\sldots$   & $0.116609\sldots$   & $-1.85190\sldots$   & $-0.111842\sldots$   & $-1.96375\sldots$   & $0.0464792\sldots$   & $-0.160426\sldots$   & $-1.40975\sldots$   & $-1.57018\sldots$   \\
    5   & $0.576137\sldots$   & $-4.23863\sldots$   & $15.8505\sldots$   & $11.6119\sldots$   & $0.117046\sldots$   & $-2.92064\sldots$   & $1.05226\sldots$   & $-1.86838\sldots$   & $0.0533164\sldots$   & $-0.818135\sldots$   & $-0.995603\sldots$   & $-1.81374\sldots$   \\
    6   & $0.534872\sldots$   & $-5.58153\sldots$   & $19.6958\sldots$   & $14.1143\sldots$   & $0.115561\sldots$   & $-4.08661\sldots$   & $2.38235\sldots$   & $-1.70426\sldots$   & $0.0568683\sldots$   & $-1.63720\sldots$   & $-0.293205\sldots$   & $-1.93041\sldots$   \\
    7   & $0.50165\sldots$   & $-6.9769\sldots$   & $23.4922\sldots$   & $16.5153\sldots$   & $0.113315\sldots$   & $-5.32460\sldots$   & $3.82394\sldots$   & $-1.50066\sldots$   & $0.0586798\sldots$   & $-2.57138\sldots$   & $0.598918\sldots$   & $-1.97246\sldots$   \\
    8   & $0.474130\sldots$   & $-8.41392\sldots$   & $27.2440\sldots$   & $18.8301\sldots$   & $0.110796\sldots$   & $-6.61868\sldots$   & $5.34600\sldots$   & $-1.27269\sldots$   & $0.0595055\sldots$   & $-3.59273\sldots$   & $1.62585\sldots$   & $-1.96689\sldots$   \\
    9   & $0.450832\sldots$   & $-9.88502\sldots$   & $30.9548\sldots$   & $21.0698\sldots$   & $0.108225\sldots$   & $-7.95803\sldots$   & $6.92912\sldots$   & $-1.02891\sldots$   & $0.0597461\sldots$   & $-4.68294\sldots$   & $2.75391\sldots$   & $-1.92903\sldots$   \\
    10   & $0.430766\sldots$   & $-11.3847\sldots$   & $34.6282\sldots$   & $23.2435\sldots$   & $0.105704\sldots$   & $-9.33488\sldots$   & $8.56031\sldots$   & $-0.774574\sldots$   & $0.0596284\sldots$   & $-5.82927\sldots$   & $3.96095\sldots$   & $-1.86831\sldots$   \\
    11   & $0.413239\sldots$   & $-12.9087\sldots$   & $38.2671\sldots$   & $25.3583\sldots$   & $0.103278\sldots$   & $-10.7435\sldots$   & $10.2304\sldots$   & $-0.513070\sldots$   & $0.0592871\sldots$   & $-7.02245\sldots$   & $5.23161\sldots$   & $-1.79084\sldots$   \\
    12   & $0.397753\sldots$   & $-14.4539\sldots$   & $41.8741\sldots$   & $27.4202\sldots$   & $0.100969\sldots$   & $-12.1794\sldots$   & $11.9327\sldots$   & $-0.246672\sldots$   & $0.0588057\sldots$   & $-8.25550\sldots$   & $6.55476\sldots$   & $-1.70074\sldots$   \\
    13   & $0.383935\sldots$   & $-16.0177\sldots$   & $45.4517\sldots$   & $29.4340\sldots$   & $0.0987797\sldots$   & $-13.6392\sldots$   & $13.6622\sldots$   & $0.0230389\sldots$   & $0.0582372\sldots$   & $-9.52303\sldots$   & $7.92208\sldots$   & $-1.60095\sldots$   \\
    14   & $0.371504\sldots$   & $-17.5979\sldots$   & $49.0018\sldots$   & $31.4039\sldots$   & $0.0967099\sldots$   & $-15.1200\sldots$   & $15.4150\sldots$   & $0.294935\sldots$   & $0.0576162\sldots$   & $-10.8207\sldots$   & $9.32717\sldots$   & $-1.49356\sldots$   \\
    15   & $0.360239\sldots$   & $-19.1928\sldots$   & $52.5264\sldots$   & $33.3335\sldots$   & $0.0947542\sldots$   & $-16.6196\sldots$   & $17.1878\sldots$   & $0.568195\sldots$   & $0.0569656\sldots$   & $-12.1451\sldots$   & $10.7650\sldots$   & $-1.38014\sldots$   \\
    16   & $0.349968\sldots$   & $-20.8010\sldots$   & $56.0269\sldots$   & $35.2259\sldots$   & $0.0929060\sldots$   & $-18.1360\sldots$   & $18.9782\sldots$   & $0.842211\sldots$   & $0.0563009\sldots$   & $-13.4934\sldots$   & $12.2315\sldots$   & $-1.26189\sldots$   \\
    17   & $0.34055\sldots$   & $-22.4213\sldots$   & $59.5049\sldots$   & $37.0836\sldots$   & $0.0911579\sldots$   & $-19.6677\sldots$   & $20.7842\sldots$   & $1.11653\sldots$   & $0.0556325\sldots$   & $-14.8632\sldots$   & $13.7235\sldots$   & $-1.13973\sldots$   \\
    18   & $0.331874\sldots$   & $-24.0525\sldots$   & $62.9616\sldots$   & $38.9091\sldots$   & $0.0895030\sldots$   & $-21.2133\sldots$   & $22.6041\sldots$   & $1.39080\sldots$   & $0.0549676\sldots$   & $-16.2524\sldots$   & $15.2380\sldots$   & $-1.01437\sldots$   \\
    19   & $0.323844\sldots$   & $-25.6939\sldots$   & $66.3983\sldots$   & $40.7044\sldots$   & $0.0879343\sldots$   & $-22.7715\sldots$   & $24.4363\sldots$   & $1.66475\sldots$   & $0.0543109\sldots$   & $-17.6594\sldots$   & $16.7730\sldots$   & $-0.886401\sldots$   \\
    20   & $0.316385\sldots$   & $-27.3446\sldots$   & $69.8160\sldots$   & $42.4715\sldots$   & $0.0864452\sldots$   & $-24.3415\sldots$   & $26.2797\sldots$   & $1.93819\sldots$   & $0.0536657\sldots$   & $-19.0827\sldots$   & $18.3264\sldots$   & $-0.756273\sldots$   \\
    21   & $0.309431\sldots$   & $-29.0039\sldots$   & $73.2157\sldots$   & $44.2118\sldots$   & $0.0850299\sldots$   & $-25.9223\sldots$   & $28.1333\sldots$   & $2.21096\sldots$   & $0.0530340\sldots$   & $-20.5210\sldots$   & $19.8966\sldots$   & $-0.624366\sldots$   \\
    22   & $0.302927\sldots$   & $-30.6712\sldots$   & $76.5983\sldots$   & $45.9271\sldots$   & $0.0836828\sldots$   & $-27.5132\sldots$   & $29.9961\sldots$   & $2.48292\sldots$   & $0.0524171\sldots$   & $-21.9732\sldots$   & $21.4822\sldots$   & $-0.490990\sldots$   \\
    23   & $0.296826\sldots$   & $-32.3460\sldots$   & $79.9645\sldots$   & $47.6185\sldots$   & $0.0823990\sldots$   & $-29.1133\sldots$   & $31.8673\sldots$   & $2.75400\sldots$   & $0.0518158\sldots$   & $-23.4383\sldots$   & $23.0819\sldots$   & $-0.356402\sldots$   \\
    24   & $0.291088\sldots$   & $-34.0278\sldots$   & $83.3151\sldots$   & $49.2874\sldots$   & $0.0811738\sldots$   & $-30.7222\sldots$   & $33.7464\sldots$   & $3.02411\sldots$   & $0.0512305\sldots$   & $-24.9153\sldots$   & $24.6945\sldots$   & $-0.220815\sldots$   \\
    25   & $0.285677\sldots$   & $-35.7161\sldots$   & $86.6509\sldots$   & $50.9348\sldots$   & $0.0800030\sldots$   & $-32.3393\sldots$   & $35.6325\sldots$   & $3.29320\sldots$   & $0.0506612\sldots$   & $-26.4037\sldots$   & $26.3193\sldots$   & $-0.0844091\sldots$   \\
    26   & $0.280564\sldots$   & $-37.4106\sldots$   & $89.9724\sldots$   & $52.5618\sldots$   & $0.0788828\sldots$   & $-33.9641\sldots$   & $37.5254\sldots$   & $3.56123\sldots$   & $0.0501078\sldots$   & $-27.9026\sldots$   & $27.9552\sldots$   & $0.0526638\sldots$   \\
    27   & $0.275723\sldots$   & $-39.1110\sldots$   & $93.2802\sldots$   & $54.1692\sldots$   & $0.0778098\sldots$   & $-35.5961\sldots$   & $39.4243\sldots$   & $3.82817\sldots$   & $0.0495701\sldots$   & $-29.4114\sldots$   & $29.6016\sldots$   & $0.190274\sldots$   \\
    28   & $0.271128\sldots$   & $-40.8168\sldots$   & $96.5749\sldots$   & $55.7581\sldots$   & $0.0767809\sldots$   & $-37.2350\sldots$   & $41.3290\sldots$   & $4.09399\sldots$   & $0.0490476\sldots$   & $-30.9295\sldots$   & $31.2578\sldots$   & $0.328313\sldots$   \\
    29   & $0.266761\sldots$   & $-42.5279\sldots$   & $99.8570\sldots$   & $57.3292\sldots$   & $0.0757931\sldots$   & $-38.8803\sldots$   & $43.2390\sldots$   & $4.35868\sldots$   & $0.0485400\sldots$   & $-32.4564\sldots$   & $32.9231\sldots$   & $0.466684\sldots$   \\
    30   & $0.262603\sldots$   & $-44.2438\sldots$   & $103.127\sldots$   & $58.8832\sldots$   & $0.0748437\sldots$   & $-40.5318\sldots$   & $45.1762\sldots$   & $4.64439\sldots$   & $0.0480469\sldots$   & $-33.9918\sldots$   & $34.5971\sldots$   & $0.605307\sldots$   \\
    31   & $0.258637\sldots$   & $-45.9645\sldots$   & $106.385\sldots$   & $60.4207\sldots$   & $0.0739304\sldots$   & $-42.1891\sldots$   & $47.0737\sldots$   & $4.88464\sldots$   & $0.0475677\sldots$   & $-35.5348\sldots$   & $36.2791\sldots$   & $0.744267\sldots$   \\
    32   & $0.254849\sldots$   & $-47.6896\sldots$   & $109.632\sldots$   & $61.9426\sldots$   & $0.0730510\sldots$   & $-43.8519\sldots$   & $48.9978\sldots$   & $5.14590\sldots$   & $0.0471020\sldots$   & $-37.0856\sldots$   & $37.9688\sldots$   & $0.883173\sldots$   \\
    33   & $0.251227\sldots$   & $-49.4190\sldots$   & $112.868\sldots$   & $63.4493\sldots$   & $0.0722034\sldots$   & $-45.5201\sldots$   & $50.9260\sldots$   & $5.40589\sldots$   & $0.0466492\sldots$   & $-38.6436\sldots$   & $39.6657\sldots$   & $1.02209\sldots$   \\
    34   & $0.247757\sldots$   & $-51.1525\sldots$   & $116.094\sldots$   & $64.9414\sldots$   & $0.0713858\sldots$   & $-47.1931\sldots$   & $52.8581\sldots$   & $5.66498\sldots$   & $0.0462090\sldots$   & $-40.2085\sldots$   & $41.3695\sldots$   & $1.16098\sldots$   \\
    35   & $0.244430\sldots$   & $-52.8899\sldots$   & $119.309\sldots$   & $66.4194\sldots$   & $0.0705964\sldots$   & $-48.8710\sldots$   & $54.7939\sldots$   & $5.92294\sldots$   & $0.0457808\sldots$   & $-41.7797\sldots$   & $43.0798\sldots$   & $1.30008\sldots$   \\
    36   & $0.241237\sldots$   & $-54.6310\sldots$   & $122.515\sldots$   & $67.8839\sldots$   & $0.0698336\sldots$   & $-50.5537\sldots$   & $56.7332\sldots$   & $6.17953\sldots$   & $0.0453641\sldots$   & $-43.3572\sldots$   & $44.7963\sldots$   & $1.43904\sldots$   \\
    \hline
  \end{tabular}
\end{sideways}
\end{table}

Values of the large $cL$ expansion parameters of the occupation
numbers for $N=2$ to 36.

\begin{table}[htbp]
  \centering
  \begin{tabular}{ccccc}
    \hline
    \hline
    $N$ & $c_0^{(0)}(N)$ &  & $c_0^{(1)}(N)$ &\\
    \hline\\
    2   & $\frac{16}{\pi ^2}$   & $1.62114\ldots$   & $\frac{64}{\pi ^2}$   & $6.48456\ldots$   \\\\
    3   & $\frac{1}{3}+\frac{35}{2 \pi ^2}$   & $2.10645\ldots$   & $8+\frac{105}{\pi ^2}$   & $18.6387\ldots$   \\\\
    4   & $-\frac{2097152}{19845 \pi ^4}+\frac{320}{9 \pi ^2}$   & $2.51766\ldots$   & $-\frac{16777216}{19845 \pi ^4}+\frac{138752}{315 \pi ^2}$   & $35.9512\ldots$   \\\\
    5   & $\frac{1}{5}+\frac{7436429}{129600 \pi ^4}+\frac{4459}{216 \pi ^2}$   & $2.88069\ldots$   & $\frac{16}{3}+\frac{7436429}{12960 \pi ^4}+\frac{249613}{540 \pi ^2}$   & $58.0593\ldots$   \\\\
    \multirow{2}{*}[-0.1cm]{6}   & $\frac{193507848058308060419981312}{12748157814913474078125 \pi ^6}$  & \multirow{2}{*}[-0.1cm]{$3.20923\ldots$}   & $\frac{774031392233232241679925248}{4249385938304491359375 \pi ^6}$  & \multirow{2}{*}[-0.1cm]{$84.6855\ldots$}   \vspace{0.2cm}\\
    & $-\frac{38494793629696}{21739843125 \pi ^4}+\frac{4144}{75 \pi ^2}$ & & $-\frac{55397943205615894528}{2589628373206875 \pi ^4}+\frac{85085248}{75075 \pi ^2}$\\\\
    7   & $\frac{1}{7}+\frac{85760621135804297813}{40663643328000000 \pi ^6}-\frac{46891706849}{317520000 \pi ^4}+\frac{79679}{3000 \pi ^2}$   & $3.51155\ldots$   & $\frac{24}{5}+\frac{85760621135804297813}{2904545952000000 \pi ^6}+\frac{1528761661843}{2716560000 \pi ^4}+\frac{641803}{875 \pi ^2}$   & $115.607\ldots$   \\\\
  \end{tabular}
\caption{Values of $c_0^{(0)}(N)$ and $c_0^{(1)}(N)$ for $N=2,3,4,5,6,7$.  Note that this Table extends Table II of \cite{ref:ffgw03}}
\label{tab:nis0majorsplit}
\end{table}

\begin{table}[htbp]
  \centering
  \begin{tabular}{ccccc}
    \hline
    \hline
    $N$ & $c_1^{(0)}(N)$ &  & $c_1^{(1)}(N)$ &\\
    \hline\\
    2   & $\frac{16}{9{\pi }^2}$                                                                                                                                           & $0.180127\ldots$   & $\frac{-832}{27{\pi }^2}$                                                                                                                                                                            & $-3.12219\ldots$   \\\\
    3   & $\frac{1}{3}$                                                                                                                                                      & $0.333333\ldots$   & $-4 - \frac{35}{2{\pi }^2}$                                                                                                                                                                          & $-5.77312\ldots$   \\\\
    4   & $\frac{-6318718976}{22325625{\pi }^4} + \frac{832}{25{\pi }^2}$                                                                                                & $0.466435\ldots$   & $\frac{-52781507411968}{7032571875{\pi }^4} + \frac{1382912}{2025{\pi }^2}$                                                                                                                        & $-7.85498\ldots$   \\\\
    5   & $\frac{1}{5} - \frac{18059899}{129600{\pi }^4} + \frac{3871}{216{\pi }^2}$                                                                                     & $0.585231\ldots$   & $\frac{28}{9} - \frac{574729727}{151200{\pi }^4} + \frac{47201}{180{\pi }^2}$                                                                                                                      & $-9.34192\ldots$   \\\\
    \multirow{2}{*}[-0.1cm]{6}   & $\frac{4458566781285863348987439874048}{315703029023206220155134375\pi^6}$ & \multirow{2}{*}[-0.1cm]{$0.693364\ldots$} & $\frac{317603131762611117568514042882856845312}{850617087107250872660154262734375\pi^6}$ & \multirow{2}{*}[-0.1cm]{$-10.2255\ldots$} \vspace{0.2cm}\\
    &  $\quad- \frac{14163619272982528}{7456766191875\pi^4} + \frac{7984}{147\pi^2}$   &   & $\quad- \frac{90755550267618998878208}{1722102868182571875\pi^4} + \frac{8618142656}{6131125\pi^2}$   &    \\\\
\end{tabular}
\caption{Values of $c_1^{(0)}(N)$ and $c_1^{(1)}(N)$ for $N=2,3,4,5,6$}
\label{tab:nis1majorsplit}
\end{table}

\begin{table}[htbp]
  \centering
  \begin{tabular}{ccccc}
\hline
\hline
$N$ & $c_2^{(0)}(N)$ &  & $c_2^{(1)}(N)$ &\\
\hline\\
2 & $\frac{16}{225{\pi }^2}$ & $0.00720506\ldots$ & $\frac{-3136}{3375{\pi
}^2}$ & $-0.0941461\ldots$ \\\\
3 & $\frac{35}{36{\pi }^2}$ &
$0.0985067\ldots$ & $\frac{-385}{12{\pi }^2}$ & $-3.25072\ldots$ \\\\
4 & $\frac{-7408644521984}{132368630625{\pi }^4} +
\frac{27584}{3675{\pi }^2}$ & $0.185917\ldots$ &
$\frac{3710310553890062336}{458657305115625{\pi }^4} -
\frac{6735912448}{7640325{\pi }^2}$ & $-6.28072\ldots$ \\\\ 
5 & $\frac{1}{5} - \frac{1062347}{127008{\pi }^4} + \frac{325}{216{\pi
}^2}$ & $0.266582\ldots$ & $- \frac{52}{9}  +
\frac{30358690219}{17781120{\pi }^4} - \frac{221909}{1080{\pi }^2}$ &
$-9.06869\ldots$ \\\\ 
\multirow{2}{*}[-0.1cm]{6} &$\frac{16076943096817340218487564310413312}{821143578489359378623504509375{\pi}^6}$  & \multirow{2}{*}[-0.1cm]{$0.34121\ldots$} &
$\frac{4194702189111033289475552785337770452189184}{2212455043565959519789061237372109375{\pi}^6} $ & \multirow{2}{*}[-0.1cm]{$-11.5824\ldots$}\vspace{0.2cm}\\
 & $- \frac{47388412779564105728}{19395048865066875{\pi }^4} +
\frac{990928}{19845{\pi }^2}$ & & $- \frac{2887827552035815814833635328}{13437568680428608340625{\pi}^4} + \frac{18544396881344}{8442559125{\pi }^2}$ & 
  \end{tabular}
\caption{Values of $c_2^{(0)}(N)$ and $c_2^{(1)}(N)$ for $N=2,3,4,5,6$}
\label{tab:nis2majorsplit}
\end{table}

\begin{table}[htbp]
  \centering
  \begin{tabular}{ccccc}
\hline
\hline
$N$ & $c_0^{(1,1)}(N)$ &  & $c_0^{(1,2)}(N)$ &\\
\hline\\
2   & $-8+\frac{64}{\pi ^2}$   & $-1.51544\ldots$   & $8$   & 8   \\\\
3   & $-16+\frac{105}{\pi ^2}$   & $-5.36128\ldots$   & $24$   & 24   \\\\
4   & $-32-\frac{16777216}{19845 \pi ^4}+\frac{2560}{9 \pi ^2}$   & $-11.8587\ldots$   & $32+\frac{16384}{105 \pi ^2}$   & $47.8100\ldots$   \\\\
5   & $-48+\frac{7436429}{12960 \pi ^4}+\frac{22295}{108 \pi ^2}$   & $-21.1931\ldots$   & $\frac{160}{3}+\frac{23023}{90 \pi ^2}$   & $79.2524\ldots$   \\\\
6   & $-72+\frac{774031392233232241679925248}{4249385938304491359375 \pi ^6}-\frac{153979174518784}{7246614375 \pi ^4}+\frac{16576}{25 \pi ^2}$   & $-33.4892\ldots$   & $72-\frac{74481467421360128}{517925674641375 \pi ^4}+\frac{7061504}{15015 \pi ^2}$   & $118.175\ldots$   \\
  \end{tabular}
\caption{Values of $c_0^{(1,1)}(N)$ and $c_0^{(1,2)}(N)$ for $N=2,3,4,5,6$}
\label{tab:nis0minorsplit}
\end{table}

\begin{table}[htbp]
  \centering
  \begin{tabular}{ccccc}
\hline
\hline
$N$ & $c_1^{(1,1)}(N)$ &  & $c_1^{(1,2)}(N)$ &\\
\hline\\
2   & $\frac{8}{3}-\frac{832}{27 \pi ^2}$   & $-0.455527\ldots$   & $-\frac{8}{3}$   & $-2.66667\ldots$   \\\\
3   & $-1-\frac{35}{2 \pi ^2}$   & $-2.77312\ldots$   & $-3$   & $-3.\ldots$   \\\\
4   & $-\frac{352}{15}-\frac{52781507411968}{7032571875 \pi ^4}+\frac{65129984}{70875 \pi ^2}$   & $-7.40762\ldots$   & $\frac{352}{15}-\frac{16728064}{70875 \pi ^2}$   & $-0.447369\ldots$   \\\\
5   & $-\frac{109}{3}-\frac{574729727}{151200 \pi ^4}+\frac{64757}{108 \pi ^2}$   & $-14.6032\ldots$   & $\frac{355}{9}-\frac{91091}{270 \pi ^2}$   & $5.26130\ldots$   \\\\
\multirow{2}{*}[-0.1cm]{6}   & $-\frac{2232}{35}+\frac{317603131762611117568514042882856845312}{850617087107250872660154262734375 \pi ^6}$   & \multirow{2}{*}[-0.1cm]{$-24.5196\ldots$}   & $\frac{2232}{35}-\frac{2698870354841096421376}{1095883643388909375 \pi ^4}$  & \multirow{2}{*}[-0.1cm]{$14.2941\ldots$}   \vspace{0.2cm}\\
 & $-\frac{40373418531338728767488}{803648005151866875 \pi ^4}+\frac{30246667072}{18393375 \pi ^2}$ & &  $-\frac{399294464}{1672125 \pi ^2}$   &\\
  \end{tabular}
\caption{Values of $c_1^{(1,1)}(N)$ and $c_1^{(1,2)}(N)$ for $N=2,3,4,5,6$}
\label{tab:nis1minorsplit}
\end{table}

\begin{table}[htbp]
  \centering
  \begin{tabular}{ccccc}
\hline
\hline
$N$ & $c_2^{(1,1)}(N)$ &  & $c_2^{(1,2)}(N)$ &\\
\hline\\
2   & $\frac{8}{15}-\frac{3136}{3375 \pi ^2}$   & $0.439187\ldots$   & $-\frac{8}{15}$   & $-0.533333\ldots$   \\\\
3   & $4-\frac{385}{12 \pi ^2}$   & $0.749279\ldots$   & $-4$   & $-4$   \\\\
4   & $\frac{2272}{105}+\frac{3710310553890062336}{458657305115625 \pi ^4}-\frac{39711774208}{38201625 \pi ^2}$   & $-0.641704\ldots$   & $-\frac{2272}{105}+\frac{6032211968}{38201625 \pi ^2}$   & $-5.63902\ldots$   \\\\
5   & $\frac{31}{3}+\frac{30358690219}{17781120 \pi ^4}-\frac{238405}{756 \pi ^2}$   & $-4.09068\ldots$   & $-\frac{145}{9}+\frac{830687}{7560 \pi ^2}$   & $-4.97802\ldots$   \\\\
\multirow{2}{*}[-0.1cm]{6}   & $-\frac{3592}{105}+\frac{4194702189111033289475552785337770452189184}{2212455043565959519789061237372109375 \pi ^6}$  & \multirow{2}{*}[-0.1cm]{$-9.82321\ldots$}   & $\frac{3592}{105}+\frac{1983596175418523964988719104}{94062980763000258384375 \pi ^4}$   & \multirow{2}{*}[-0.1cm]{$-1.75923\ldots$}   \vspace{0.2cm}\\
 & $-\frac{98659506842974376305885184}{418057692280001148375 \pi ^4}+\frac{13193444962112}{2814186375 \pi ^2}$ & & $-\frac{21035938004992}{8442559125 \pi ^2}$
  \end{tabular}
\caption{Values of $c_2^{(1,1)}(N)$ and $c_2^{(1,2)}(N)$ for $N=2,3,4,5,6$}
\label{tab:nis2minorsplit}
\end{table}

\end{document}